\documentclass[journal=jacsat,manuscript=article]{achemso}

\usepackage[version=3]{mhchem} 

 \usepackage{threeparttable}
\usepackage{float}
\restylefloat{table}
\usepackage{afterpage}

\usepackage{xr}
\makeatletter
\newcommand*{\addFileDependency}[1]{
  \typeout{(#1)}
  \@addtofilelist{#1}
  \IfFileExists{#1}{}{\typeout{No file #1.}}
}
\makeatother

\newcommand*{\myexternaldocument}[1]{
    \externaldocument{#1}
    \addFileDependency{#1.tex}
    \addFileDependency{#1.aux}
}

\myexternaldocument{supp_info}

 \usepackage{graphicx}
\usepackage{dcolumn}
\usepackage{bm}
 \usepackage{braket} 

\graphicspath{ {images/} }

\title{Excitonic effects in X-ray absorption spectra of fluoride salts and their surfaces }%
\author{Ana Sanz-Matias}
\affiliation{%
 Joint Center for Energy Storage Research, Lawrence Berkeley National Laboratory, Berkeley, CA 94720, USA
}%
\alsoaffiliation{The Molecular Foundry, Lawrence Berkeley National Laboratory, Berkeley, CA 94720, USA}

\author{Subhayan Roychoudhury}
\affiliation{The Molecular Foundry, Lawrence Berkeley National Laboratory, Berkeley, CA 94720, USA}
\author{Xuefei Feng}
\affiliation{%
 Joint Center for Energy Storage Research, Lawrence Berkeley National Laboratory, Berkeley, CA 94720, USA
}%
\alsoaffiliation{%
 Advanced Light Source, Lawrence Berkeley National Laboratory, Berkeley, CA 94720, USA
}%
\author{ Feipeng Yang}
\affiliation{%
 Joint Center for Energy Storage Research, Lawrence Berkeley National Laboratory, Berkeley, CA 94720, USA
}%
\alsoaffiliation{%
 Advanced Light Source, Lawrence Berkeley National Laboratory, Berkeley, CA 94720, USA
}%
\author{Li Cheng Kao}
\affiliation{%
 Joint Center for Energy Storage Research, Lawrence Berkeley National Laboratory, Berkeley, CA 94720, USA
}%
\alsoaffiliation{%
 Advanced Light Source, Lawrence Berkeley National Laboratory, Berkeley, CA 94720, USA
}%

\author{Kevin R. Zavadil}
\affiliation{Joint Center for Energy Storage Research, Lemont, Illinois 60439, United States}
\alsoaffiliation{Material, Physical and Chemical Sciences Center, Sandia National Laboratories, Albuquerque, New Mexico 87185, United States}

\author{Jinghua Guo}
\affiliation{%
 Joint Center for Energy Storage Research, Lawrence Berkeley National Laboratory, Berkeley, CA 94720, USA
}%
\alsoaffiliation{%
 Advanced Light Source, Lawrence Berkeley National Laboratory, Berkeley, CA 94720, USA
}%
\author{David Prendergast}%
\affiliation{%
 Joint Center for Energy Storage Research, Lawrence Berkeley National Laboratory, Berkeley, CA 94720, USA
}%
\alsoaffiliation{The Molecular Foundry, Lawrence Berkeley National Laboratory, Berkeley, CA 94720, USA}
 \email{dgprendergast@lbl.gov}
 
\begin{document}

\begin{tocentry}
\includegraphics[width=0.75\linewidth]{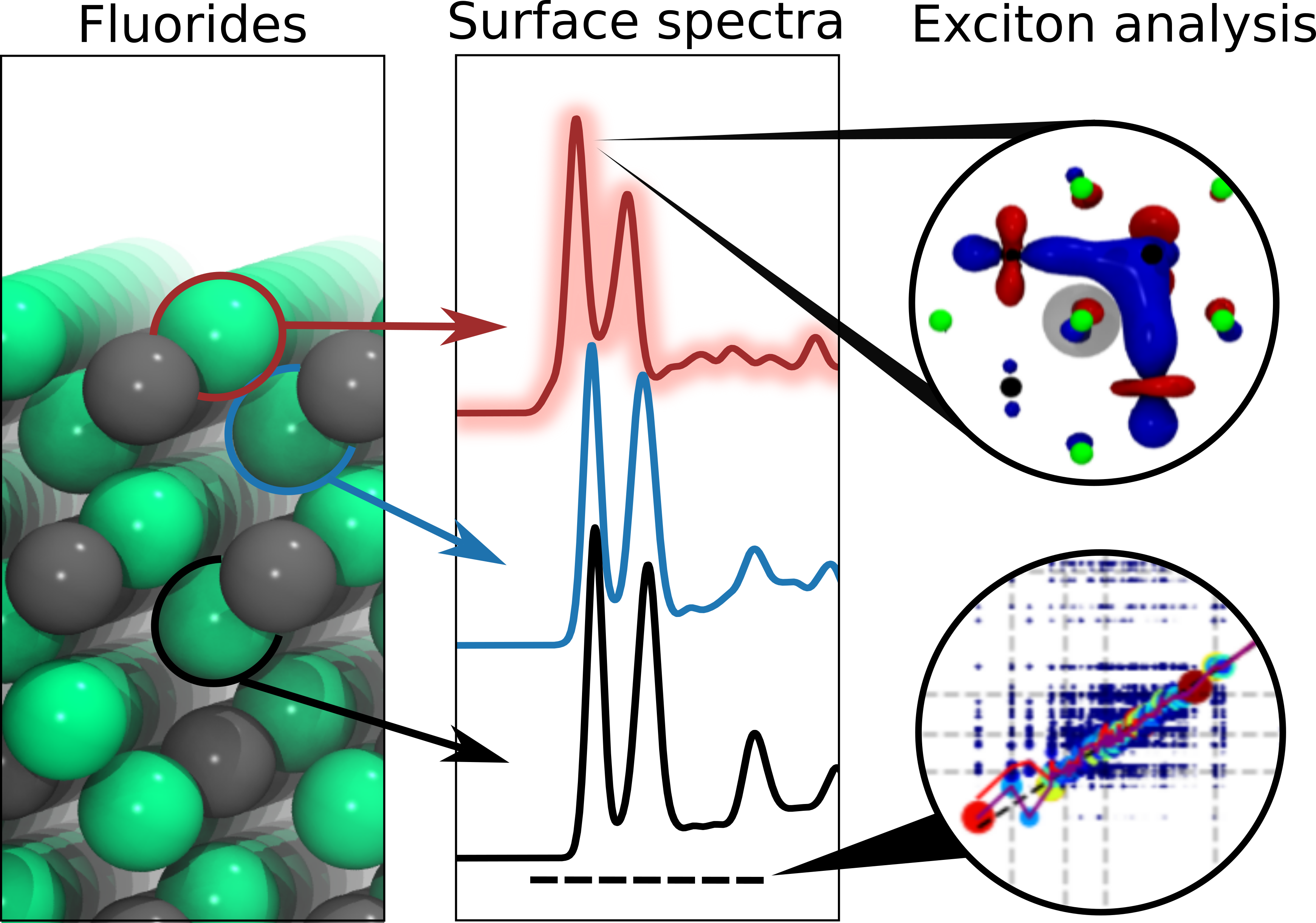}
\end{tocentry}

\begin{abstract}
Given their natural abundance and thermodynamic stability, fluoride salts may appear as evolving components of electrochemical interfaces in Li-ion batteries and emergent multivalent ion cells. This is due to the practice of employing electrolytes with fluorine-containing species (salt, solvent, or additives) that electrochemically decompose and deposit on the electrodes. Operando X-ray absorption spectroscopy (XAS) can probe the electrode-electrolyte interface with single-digit nanometer depth resolution and offers a wealth of insight into the evolution and Coulombic efficiency or degradation of prototype cells, provided that the spectra can be reliably interpreted in terms of local oxidation state, atomic coordination, and electronic structure about the excited atoms. To this end, we explore fluorine K-edge XAS of mono- (Li, Na, K) and di-valent (Mg, Ca, Zn) fluoride salts from a theoretical standpoint and discover a surprising level of detailed electronic structure information about these materials, despite the relatively predictable oxidation state and ionicity of the fluoride anion and the metal cation. Utilizing a recently developed many-body approach based on the $\Delta$SCF method, we calculate the XAS using density functional theory and experimental spectral profiles are well reproduced, despite some experimental discrepancies in energy alignment within the literature, which we can correct for in our simulations. We outline a general methodology to explain shifts in the main XAS peak energies in terms of a simple exciton model and explain line-shape differences resulting from mixing of core-excited states with metal $d$ character (for K and Ca specifically). Given ultimate applications to evolving interfaces, some understanding of the role of surfaces and their terminations in defining new spectral features is provided to indicate the sensitivity of such measurements to changes in interfacial chemistry.
\end{abstract}

\section*{Introduction}

Fluorinated species continue to play a significant role in the development of rechargeable batteries because they are considered to be consistently beneficial for safety and electrochemical performance, attributed in part to the formation of effective interphases at anode and cathode rich in native fluorinated salts.~\cite{vonaspern2019} Known as solid and cathode electrolyte interphases (SEI/CEI), respectively, these are grown through the decomposition of additives or fluorinated electrolytes, or  through the synthesis of an artificial SEI.~\cite{LiuLiu2020,YU2020}

The contribution of fluorine-rich interphases to  additional stability has been extensively studied in recent years: in Li-anode batteries, $ex$-$situ$ SEIs enriched with LiF~\cite{fan2018, cui2020}  have been used to prevent dendrite growth in Li anodes;  $in$-$situ$ fluorine-rich SEIs are critical to achieve high Coulombic efficiencies.~\cite{He2020} Following the examples for Li, fluorine-rich interphases have been recently used in Na batteries, with NaF-rich interphases.~\cite{liuxuyang2021} In the emergent field of mutivalent cation-based chemistries, this additional stability has facilitated the very recent development of reversible batteries using multivalent ions (Mg, Ca, Zn). MgF$_2$ has been identified as a suitable protective layer for anode and cathode in magnesium-based batteries;~\cite{chen2019,li2019} the dynamic character of Ca interfacial behavior has been studied by comparing pre-formed CaF$_2$ interfaces with the native CaF$_2$-containing SEI;~\cite{Melemed_2020} and ZnF$_2$-rich SEIs have been detected in dendrite-free, reversible Zn setups.~\cite{Qiu2019,Cao2021} In addition, fluoride salts such as ZnF$_2$ have been used as anodes in fluoride-ion batteries.~\cite{Nowroozi2020} Yet, the complexity of these interfacial transformations is only beginning to be understood.~\cite{gong2021}

\bigskip

One characterization technique that has the potential to probe  and  reveal chemical details of such buried interphases is X-ray absorption spectroscopy (XAS). XAS can be measured in bulk-sensitive modes, such as absorption deduced by transmission through a thin film or, more commonly, through measurement of integrated secondary photon emission -- so-called Total Fluorescence Yield (TFY) -- with such fluorescence emerging from depths of microns below a sample surface. XAS can also be measured in surface-sensitive modes through collection of secondary electron emission (typically Auger electrons) -- so-called Total Electron Yield (TEY) -- with a depth profile of typically less than 5 nm defined by the inelastic mean free path of electrons. In combination, these various XAS measurement modalities enable advanced in-situ/operando characterization techniques to provide detailed insight on complex environments within electrochemical systems such as batteries.~\cite{Kao2020} Knowing the spectral signatures of fluorinated salts is key for interphase characterization and for tracking the decomposition of fluorinated additives in the electrolyte or intentionally deposited on an electrode. 
Fluorine is the most electronegative element and typically exists as singly-charged fluoride anions (satisfying the octet rule) in minerals. Despite this, F K-edge spectra are surprisingly rich. For example, the variability of the F K edge has been proposed to distinguish uranium fluoride from other compounds;~\cite{WARD2017}  and the sensitivity of the F K edge in LiF, in particular, has been studied experimentally and computationally, both in terms of finite temperature effects and the presence of optically-induced decomposition products.~\cite{hamalainen2002, hudson1994,SCHWARTZ2017}

Many-body perturbation theory, solving the Bethe-Salpeter equation,~\cite{SHIRLEY20051187} has previously been applied to accurately reproduce the experimental spectral line shape of LiF.~\cite{SCHWARTZ2017} 
Additionally, and with less computational expense, density functional theory (DFT) calculations have also seen widespread success in reproducing X-ray absorption spectra of materials and molecules, through explicit modeling of core-excited states using the so-called core-hole (CH) approach, wherein the core-excited electron density (and by extension, the self-consistent field) is constructed by vacating a specific core orbital .~\cite{shirley2021} Recently, these CH methods have advanced through inclusion of many-body contributions to XAS intensities by considering an approximation to the ground and core-excited many-body states based on single determinants comprising the occupied Kohn-Sham orbitals from their respective self-consistent fields (or electron densities), dubbed MBXAS~\cite{liang2018, liang2019}. Previously, CH methods have calculated dipole transition matrix elements using an effective single-particle approximation, only involving the core orbital of the initial (ground) state and the orbital accessed in the excited state, derived from the CH self-consistent field. The MBXAS methodology has provided improved agreement with XAS measurements, especially in the pre-edge features of transition metal oxides at the O K edge. We intend to validate the MBXAS methodology for the calculation of the F K edge, focusing on fluoride salts that may be relevant to electrochemical contexts. We rely mostly on previously measured XAS of fluoride salts, due to their sharp spectral features, and relegate discussion of more covalent flourinated molecules, whose spectra are less distinct, to future work. 
We also explore the impact of finite temperature effects on the spectra, inspired by our previous work on similar compounds at the Li K edge~\cite{pascal_2014_li} which used the single-particle CH approach. 

\bigskip
In this work, we study LiF, KF, NaF, MgF$_2$, CaF$_2$ and ZnF$_2$, which are the most stable fluorides of common or promising cations in battery research and constitute an adequate data set within the literature. 
The results are organized as follows: First, we discuss the agreement between the calculated and experimental F K-edge spectra of the selected fluorides. We highlight discrepancies in the energy calibration of previous measurements, which complicate our efforts to establish a predictive theoretical approach for interpreting measurements of indeterminate structures or chemical contexts for fluorine. We provide estimates of the impact of finite temperature based on molecular dynamics sampling. We reveal the electronic-structure origin of the dominant spectral features, with respect to local fluorine character and contributions from neighboring atomic orbitals. We explain differences in main peak energies across this series by reference to a simple exciton model, breaking the excitation energy into its occupied core and unoccupied valence orbital contributions with renormalization provided by an estimate of the associated exciton binding energy. Finally, we explore the surface spectral contributions of a select subset (LiF, MgF$_2$, and CaF$_2$) to reveal unique spectral signatures that should be evident in studies of open surfaces and evolving interfaces relevant to electrochemistry.

\section{Methodology}
\label{sec:comp_meth}
\subsection{Crystalline fluoride salt modelling}

For each of the fluorides, the most stable room-temperature phase was selected: rock salt for LiF, NaF and KF, rutile for MgF$_2$ and ZnF$_2$, and fluorite for CaF$_2$. The coordination around the fluoride anion differs depending on the phase, from 6 to 3 to 4, respectively, as shown in Table~\ref{tab:latpars}. The bulk solid and surface atomic structure were optimized using density functional theory (DFT) with the Quantum ESPRESSO package.~\cite{Giannozzi_2009} The PBE generalized-gradient exchange-correlation functional~\cite{perdew1996} was used to describe the electronic structure within the plane-wave pseudopotential framework. Ultrasoft pseudopotentials~\cite{vanderbilt1990} represent valence electron scattering from the core electrons, and a plane-wave kinetic energy cutoff of 25 (200) Ry was used to describe the valence orbitals (electron density). A 10$\times$10$\times$10 k-point grid  was found to be sufficient to sample the Brillouin zone in all cases. The obtained lattice parameters are in reasonable agreement with experimental values (Table~\ref{tab:latpars}), with an overestimation ranging from 0.8\% (LiF) to 2.9\% (NaF) as expected for the PBE functional.~\cite{ZhangTS2018}

 Ab-initio molecular dynamics simulations on 2$\times$2$\times$2 supercells were carried out with VASP. These $\Gamma$-point calculations were also performed using the PBE exchange-correlation functional.
 First, supercells were equilibrated for (at least) 10 ps in the NVT ensemble, with a timestep of 1 fs at a temperature of 298 K,  using a Nos\'e-Hoover thermostat,\cite{nose1984, hoover1985} followed by another 10 ps for sampling. For each solid, five or six time-separated snapshots sampled from the final 5 ps of the production simulation were used to compute the temperature-averaged spectra (totalling more than 200 individual atomic spectra in each case, sampling all atoms of the same element in each snapshot).~\cite{pascal_2014_li}

Surfaces were generated from the bulk optimized structures using the Atomic Simulation Environment.~\cite{ase-paper} Slabs of more than 5 layers were used in order to obtain a well converged surface relaxation~\cite{Puchin2001} as well as a bulk-like character in the center of the slab.  The positions of the atoms in the central layer of the slab were frozen during optimization so as to mimic the bulk structure, with minimal relaxation in the case of LiF and CaF$_2$ and a small degree of distortion in the outermost MgF$_2$ layers. 
While there is only one possible termination for (001) slabs, that is not the case for fluorite (111) slabs. 
Fluorite-type (111) slabs that conserve the MX$_2$ stoichiometry can have two types of surface terminations.~\cite{Tasker1979} If the plane is cleaved in between MX$_2$ layers the surface is Type 2: F-terminated (and equivalent) on both faces of the slab. While there is no overall slab dipole, there is a local surface dipole on the  outermost edge of the surface because the slab layers are charged. This is the most stable cleavage.Type 3 slabs are obtained if, on the other hand,  the cleavage is done between M-X planes: termination on both faces of the slab is different, with one side being cation-terminated and the other doubly-anion terminated. This side is unstable and there is an overall dipole in the slab. We focus here on the Type 2 termination.

\begin{table*}[]
    \centering
    \begin{threeparttable}
   \resizebox{\textwidth}{!}{ \begin{tabular}{ccccccccccccc}
         Formula &Structure & Symmetry &  CN$_{\textnormal F}$ & Local F symmetry  & a$_0$(\AA) & c$_0$ (\AA); u & a (Exp.) & c (Exp.) &   Supercell & N & \\
         \hline
         LiF & rock salt & Fm$\bar{3}$m &6 & octahedral & 4.022 & - & 3.99\tnote{$^a$}& - &  3x3x3 & 216  \\
         NaF & rock salt & Fm$\bar{3}$m &  6 & octahedral & 4.704 &-& 4.57\tnote{$^a$}& - & 3x3x3 & 216  \\
         KF & rock salt & Fm$\bar{3}$m & 6 & octahedral & 5.433 & - & 5.347\tnote{$^a$}& - &2x2x2 & 64 \\
         MgF$_2$ & rutile & P4$_2$/mnm & 3 & trigonal &4.683& 3.086;0.302 &4.621&3.052\tnote{$^b$}& 3x3x3 & 162  \\
         CaF$_2$ & fluorite & Fm$\bar{3}$m & 4 & tetrahedral & 5.511&-&5.464\tnote{$^c$} &-&2x2x2 & 96  \\
         ZnF$_2$ & rutile & P4$_2$/mnm & 3 & trigonal & 4.782&3.171;0.304 &4.703&3.133\tnote{$^d$}&3x3x3 & 162 \\
        \hline
    \end{tabular}}
    \begin{tablenotes}\footnotesize
\item[a] From Ref.~\cite{wyckoff1931}
\item[b] From Ref.~\cite{haines2001}
\item[c] From Ref.~\cite{2007handbook}
\item[d] From Ref.~\cite{OToole2001}
\end{tablenotes}
\end{threeparttable}

    \caption{Structural details of the fluoride salts considered in this study: chemical formula, crystal structure, space group (in Hermann-Mauguin notation), coordination number (CN$_{\textnormal F}$) and local symmetry around the fluorine atom, and optimized lattice parameters (\AA), as well as experimental lattice parameters from Refs.~\cite{wyckoff1931,haines2001,2007handbook,OToole2001}. For the rutile phases we include the internal parameter $u$ that defines the F positions as $\pm$ ($u$, $u$, 0; 1/2 + $u$, 1/2 $-$ $u$, 1/2). In all cases $\alpha$ = $\beta$ = $\gamma$ = 90. The size and number of atoms (N) of the supercell used in the XAS calculations is given in the last two columns.}
    \label{tab:latpars}
\end{table*}

\subsection{X-ray absorption simulations}

X-ray absorption spectra were calculated using the MBXAS methodology.~\cite{liang2018,liang2019} This is an extension of the core-hole (CH) approaches which model core-excited electronic structure by placing constraints on the orbital occupancy within Kohn-Sham (KS) DFT, emptying a specific core-orbital and converging an excited state self-consistent field by contrast with the ground state (so-called $\Delta$SCF). Whereas CH approaches compute X-ray transition matrix elements within a single-particle approximation (involving only the core-orbital and a single final-state orbital), the MBXAS approach includes all occupied orbitals by approximating each many-body state as a single Slater determinant constructed using  the occupied orbitals from the relevant self-consistent field. The Slater determinant of the final (initial) state is built using the KS orbitals obtained with(out) a core-hole.

\bigskip

We model the core-excited states within a real-space impurity model, where a core-excited atom (here represented by a modified pseudopotential) is embedded within a supercell of the given crystal. Supercells were generated from the optimized bulk structures and are necessary, within periodic calculations, in order to reduce spurious interactions between periodic images of core-excited atoms. Supercells affording isotropic core-hole separations of at least 1~nm  usually provide enough screening to prevent these effects. Supercell sizes (see Table~\ref{tab:latpars}) were selected on the basis of convergence tests  to obtain adequately converged spectra (Fig. \ref{fig:si_super}).  However,  due to prohibitive computational cost, all temperature-averaged spectra were computed using a 2x2x2 supercell. 

\bigskip

In these pseudopotential calculations, we have no explicit inclusion of core electron orbitals nor their contribution to the total electronic energy and to the nodal structure of valence orbitals comprising atomic orbital contributions with higher principal quantum number. With respect to calculated transition matrix elements, we used the core orbital obtained on an atom-centered log-radial grid during the pseudopotential generation for the excited atom and employed the projector-augmented wave (PAW) formalism of Bl\"ochl~\cite{blochl1994} to augment valence pseudo-orbitals within the frozen-core approximation.~\cite{taillefumier2002} For the estimation of the excitation energies, we adopted the following procedure: (1) CH orbitals were generated within the full-core-hole (FCH) approximation, which includes the core-hole but neglects the excited electron (analogous to modeling core photoionization); (2) Relative excitation energies based on different CH final state orbitals were computed using the KS eigenvalue difference between specific final state orbitals accessible in the unoccupied subspace and the first available orbital above the Fermi level (effectively the conduction band minimum of the CH system); (3) Relative many-body energy alignment is provided using the SCF energy that combines the CH state with placing the excited electron in this first available orbital, referred to as the excited-electron and core-hole (XCH) approximation.~\cite{prendergast2006}. Theoretical alignment of a specific XCH state excitation energy (for a given fluoride salt in this case) is made with respect to a theoretical reference -- the isolated atom (F in this case). This permits us to provide consistent alignment between different materials and/or chemical environments and different numerical approximations, as energetic comparisons are made with respect to the same numerical parameters including the same supercell dimensions for the isolated atom calculations.~\cite{ENGLAND2011} An additional (global) rigid empirical shift needs to be added in order to align the calculated spectrum with a chosen experimental standard. This empirical shift is typically specific to a given study, due to differences in  pseudopotential approximation. Due to the excellent agreement between different experiments and sharpness of its spectral features, we have used the CaF$_{2}$ F K-edge spectrum of Oizumi \textit{et al.}~\cite{oizumi1985} and Gao \textit{et al.}~\cite{gao1993} to obtain the specific alignment in this study. 

In order to simulate the spectra of an interface in a TEY-detection context, we performed a weighted linear combination of calculated bulk and surface spectra taking into account that the inelastic mean free path (IMFP) of electrons in this energy range ($\sim$~700~eV) for some fluorides has been measured to be 1.8~nm.~\cite{tanuma1991, Flores-Mancera2020, boutboul1996} The rate of electrons exiting the slab was assumed to exponentially decay with depth into the slab relative to the IMFP. Thus, the weight of each contributing spectrum was computed using the distance of the contributing atom from the surface (see Table \ref{tab:siweights_interface}).

\subsection{Experimental details}

The F K-edge X-ray absorption spectra (XAS) of KF, CaF$_2$ and ZnF$_2$ were measured at Beamline 7.3.1 at the Advanced Light Source (ALS), Lawrence Berkeley National Laboratory (LBNL), USA. This is a bending-magnet beamline with a photon energy range from 250 eV to 1650 eV. The base pressure of the main chamber is better than $1 \times 10^{-9}$~mbar. The total electron yield (TEY) signal was obtained by monitoring the sample drain current. All the powder samples were mounted on the sample holder using a carbon tape. To further increase the TEY signal intensity, silver paste was used to connect the surface of these samples to the metallic sample holder in order to reduce the charging effect.

\section{Results and Discussion}

\begin{figure*}[ht]
\centering
\includegraphics[width=1.2\linewidth]{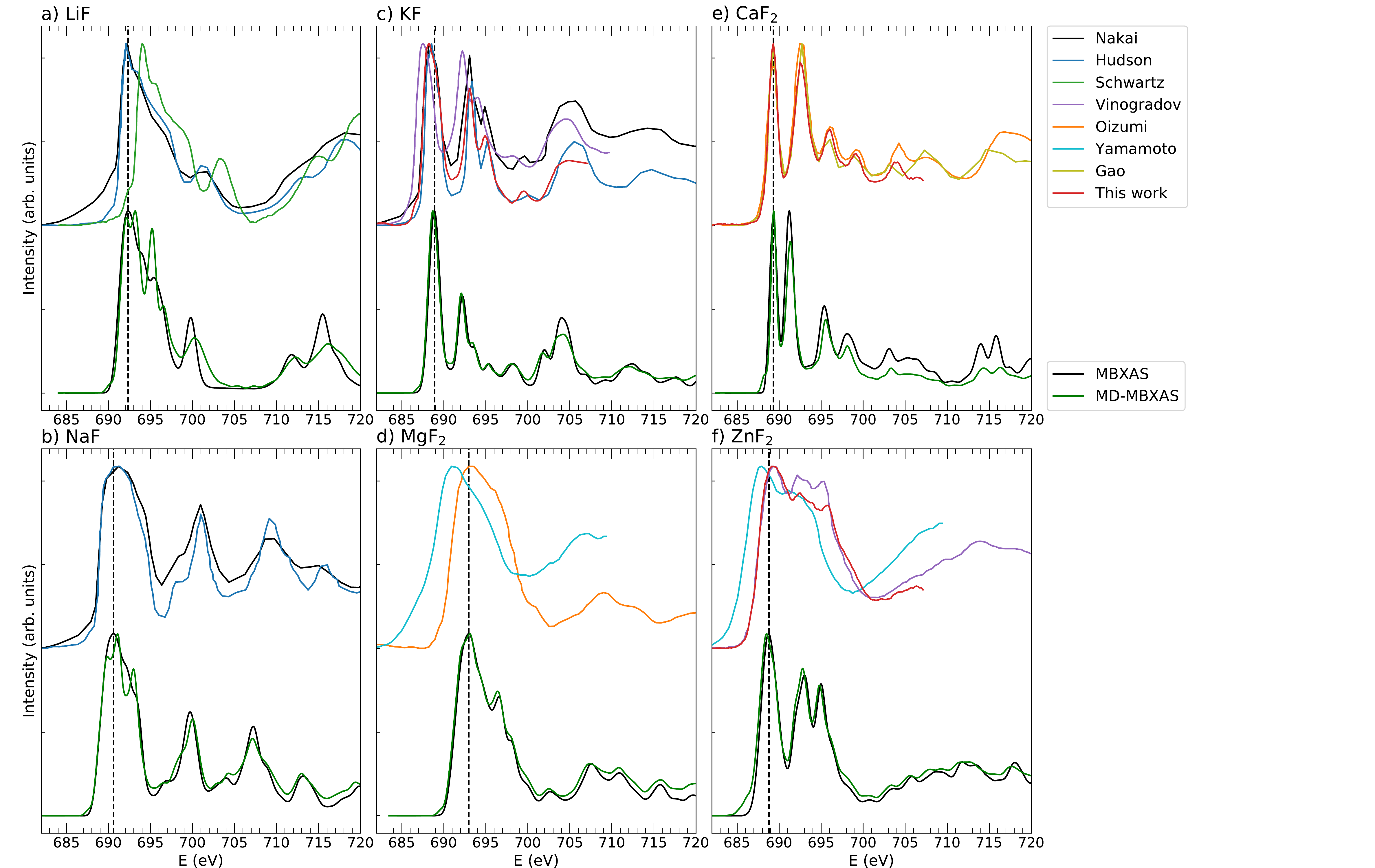}
\caption{Calculated crystal (black) and temperature-averaged (green) spectra of reference solids are shown in the bottom part of each subplot, with their corresponding experimental spectra on the top part: LiF (a),~\cite{nakai1986,hudson1994,SCHWARTZ2017} NaF (b),~\cite{nakai1986,hudson1994} KF (c),~\cite{nakai1986,   hudson1994,  vinogradov2005} MgF$_2$ (d),~\cite{oizumi1985, Yamamoto2004} CaF$_2$ (e),~\cite{oizumi1985,gao1993} and ZnF$_2$ (f).~\cite{Yamamoto2004,vinogradov2005}. Each set of experiments is shown in a different color. Note that the only true bulk absorption measurement is that of Oizumi et al.~\cite{oizumi1985} (orange), while all others use some form of surface-sensitive electron yield. The main edge peaks are marked with vertical dashed lines to ease the comparison in alignment between theory and experiment.   }
\label{fig:solid_spec}
\end{figure*}

The calculated F K-edge XAS spectra of all the solid fluorides considered here are shown in Fig.~\ref{fig:solid_spec}, together with their corresponding experimental spectra, acquired according to the Experimental Details and also collected from the literature.  The spectra are characterized by the position of the main edge that determines the energy alignment between them and the spectral line-shape, so the discussion is structured accordingly.

\subsection{Energy alignment}

\paragraph*{Absolute alignment}
At first glance, it is clear that while spectral line shapes for a given salt agree approximately between different measurements (and with our calculations) their energy alignment can vary significantly -- from essentially 0~eV for CaF$_2$ up to $\sim$1.8~eV for ZnF$_2$.~\cite{Yamamoto2004}
This experimental variation is larger than the experimental energy resolution, which ranges up to 0.4~eV~\cite{oizumi1985,gao1993}. In the case of LiF, for instance, there seem to be two types of alignments in the literature, with the main edge peak either at $\sim$~692.0~\cite{hudson1994,nakai1986} or $\sim$~694~eV.~\cite{SCHWARTZ2017,hamalainen2002} However, the latter data are aligned to a non-resonant inelastic X-ray scattering experiment with an energy resolution of only 1.9~eV, and so, perhaps, we could realign them to the former higher resolution data -- the line shapes agree very well. However, in general, this variability makes it challenging for us to choose a specific spectrum as the baseline for our own (theoretical) energy calibration for the F K edge. 

As mentioned in the Methodology section, due to the use of pseudopotentials, our computational approach requires that a \textit{single} rigid shift is added to \textit{all} calculated spectra before comparison with measured spectra. We have selected a specific CaF$_2$ F K-edge measurement by Oizumi \textit{et al.}~\cite{oizumi1985} as the reference to define the rigid shift, due to the sharpness of its spectral features and the excellent agreement with both Gao \textit{et al.}~\cite{gao1993} and our own experiment. Note that Oizumi's measurement~\cite{oizumi1985} is the only true absorption measurement of all the cited experiments, since it was measured based on X-ray transmission through a thin sample, while the others were measured using some form of surface sensitive electron yield. This also minimizes any distinct surface contributions to this measurement and makes it a genuine bulk standard for comparison with our calculations (see below for a discussion of surface contributions). 
As a result of this choice, the energy alignment of the calculated spectra are in better agreement with certain sets of experimental spectra than with others. For example, data from Nakai \textit{et al.}~\cite{nakai1986} and Hudson \textit{et al}.~\cite{hudson1994} appear closer to our simulations than those of Yamamoto \textit{et al.}~\cite{Yamamoto2004}. 

A consistent spectral energy alignment is vital for combining a predictive computational approach with difficult to interpret measurements. For example, for measurements on systems of unknown or ambiguous composition, for which the atomic or electronic structure is unclear, we can only test hypotheses if our computational predictions correspond to the same energy scale as in the experiment. Furthermore, if the measurement is made on a sample with a mixture of materials or molecular components, and/or containing various defects or impurities, then deducing this composition based on calculations alone will be extremely challenging without a reliable spectral alignment. Therefore, we emphasize the importance of \emph{well-calibrated references} for experimental spectral alignment -- particularly for future measurements of mixed phases, such as \textit{in situ/operando} studies of interphases.

\bigskip
\paragraph*{Relative alignment}

If one assumes consistent energy alignment within a given experimental report, then experiments that examine several well-defined compounds allow us to compare the \textit{relative} energy alignment between the main-edge peaks of spectra of different solids (e.g., LiF, NaF and KF in Refs.~\cite{nakai1986} and~\cite{hudson1994}). In this way, a ``distance'' matrix can be generated, in which the solids can be compared pairwise, as shown in the upper section of Table~\ref{tab:shifts}.   

The differences between the experimental and calculated relative alignment, shown in the bottom part of Table~\ref{tab:shifts}, are generally below 1~eV. However, based on previous experience~\cite{ENGLAND2011, pascal_2014_li}  we would expect agreement within 0.1-0.2~eV. The differences could be ascribed either to weaknesses within our theoretical approach (for example, our choice of density functional)~\cite{Roychoudhury2021} or to the aforementioned lack of consistency between experiments. Note that slight changes in peak energy may happen also depending on the choice of broadening parameter for the calculated spectra. However, for the moment, we can restate the excellent agreement with respect to spectral line shape that is evident in Fig.~\ref{fig:solid_spec} and look forward to future measurements that provide accompanying reference data for alignment.

\begin{table*}[]
    \centering
    \begin{threeparttable}

\resizebox{\textwidth}{!}{\begin{tabular}{l|ccc|ccc|ccc|cc|cc|ccc}

{} &   LiF & & &   NaF & & &    KF &   & & CaF$_2$ & &   MgF$_2$ & &   ZnF$_2$  & \\
\hline

LiF  &  0.00  \\
NaF  &-0.90  &\textit{-0.98$^a$}& \textit{-0.87$^d$ }&  0.00 \\
KF   &-3.53   &\textit{-3.94$^a$}&\textit{-3.66$^d$ }& -2.33 &\textit{-2.96$^a$}&\textit{ -2.79$^d$}&  0.00  \\
CaF$_2$ &-2.91  &            &                &  -2.01 &&  &   0.32&\textit{1.00 } &&0.00   \\
MgF$_2$ & 0.5   &            &                 & 1.40  & & & 3.73  &  &  & 3.41   & \textit{ 3.78$^e$} &  0.00 \\
ZnF$_2$ &   -3.65&           &   &  -2.75 && &   -0.42 &\textit{-1.74$^c$} & \textit{0.59} &  -0.74 & \textit{-0.41} &-4.15 & \textit{-3.02$^b$} &  0.00 \\
\hline

LiF  &  0.00  \\
NaF  && -0.08$^a$ &0.03$^d$& 0.00 \\
KF   && -0.41$^a$ &-0.13$^d$ &  &  -0.63$^a$ &  -0.46$^d$&0.00  \\
CaF$_2$ &  &   &   & &&&&0.68 && 0.00    \\
MgF$_2$ &   &   & &&& &&&&& -0.37$^e$ &    0.00 \\
ZnF$_2$ &  & & &&& &&-1.32$^c$  &1.01   &&0.33  & & 1.13$^b$   &  0.00 \\
\hline
\end{tabular}}

    \begin{tablenotes}\footnotesize
\item[a] From Ref.~\cite{nakai1986}
\item[b] From Ref.~\cite{Yamamoto2004}
\item[c] From Ref.~\cite{vinogradov2005}
\item[d] From Red.~\cite{hudson1994}
\item[e] From Ref.~\cite{oizumi1985}
\end{tablenotes}
\end{threeparttable}

    \caption{The top of the table contains the relative shifts (in eV, referenced to LiF) between the calculated spectra of LiF, NaF, KF, CaF$_2$, MgF$_2$ and ZnF$_2$, followed by experimental values of said shifts, shown in italics. The bottom of the table shows the differences between the calculated and the experimental relative shifts, with their corresponding literature reference. Values without references correspond to this work.}
    \label{tab:shifts}
\end{table*}

\paragraph*{Origins of relative alignment}

\begin{figure*}[htp]
\centering
\includegraphics[width=0.7\linewidth]{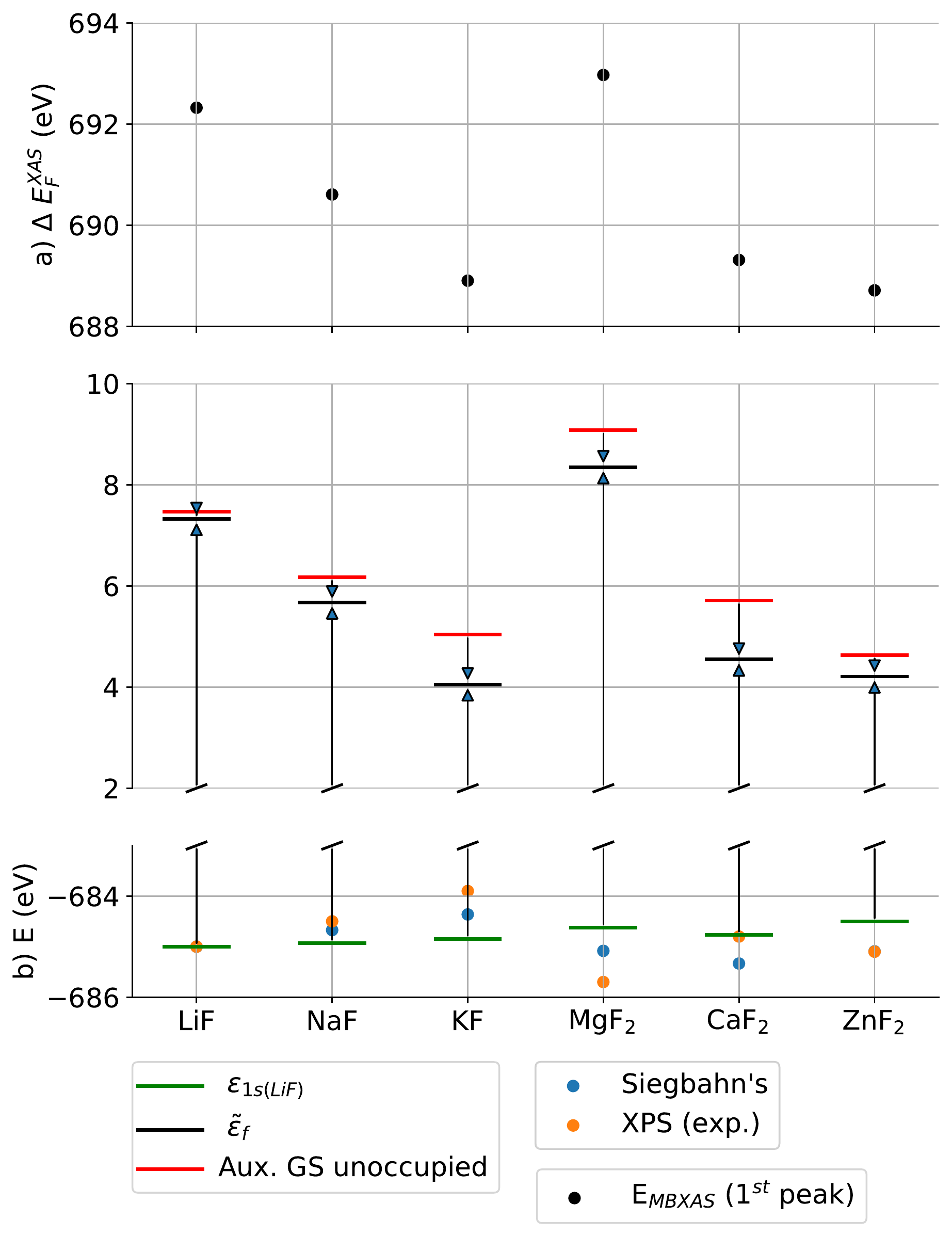}
\caption{Scheme detailing the origins of relative alignment between the first peaks (a) in the XAS spectra using the simple exciton model (b). For each solid, we selected a final state with the largest oscillator strength at the energy of the main peak, and the corresponding energy shift in the 1$s$ electron (E$_{1s}$, green line), the resonant excitation energy that leads to the final state (black line) from an auxiliary ground-state unoccupied energy level (red line). The exciton binding energy (downward arrow) connects the  final state with the auxiliary GS energy level.  Additionally, the electrostatic effects on the core hole according to Siegbahn's expression (blue dots) are shown together with the experimental XPS shift (orange dots).~\cite{Moulder1992HandbookOX, wagner1991} All energies are in electronvolts (eV). }
\label{fig:shifts}
\end{figure*}

However, the relative energy alignment between XAS spectral features of different materials does have physical origins. Red-shifts in the main-edge peaks down a given group of the periodic table (eg., LiF, NaF, KF or MgF$_2$, CaF$_2$) are perhaps more clearly recognized when the relative energy of these peaks is plotted, as in Fig.~\ref{fig:shifts}~(a).  

This raises a point that is common to many cases of spectral interpretation and the goal of understanding spectral changes correlated with changes in material or chemical context. Unlike certain other spectroscopies where spectral features at particular energies may reliably correspond to a given system response (e.g., specific molecular vibrational modes), X-ray spectral peaks are not always easily associated with certain ground state electronic structure. The core excitation can dramatically reorganize and reorder the electronic structure and, especially in condensed phase contexts, mix together contributions from many ground state electronic orbitals to define the final state energy and intensity. 

We consider breaking down spectral differences based on a simplified exciton model. For a given X-ray absorption peak, we express the resonant X-ray absorption energy associated with a specific final state as $\Delta E_F^{\rm XAS}=E_F - E_{\rm GS}$. We consider this final state energy resulting from the effective excitation energies for the core hole 
($\epsilon_{1s}$) and the excited electron ($\epsilon_f$), and the renormalization of their effective single-particle energy difference due to exciton binding ($E_b^F$), expressed as:
\begin{equation}
    \Delta E_F^{\rm XAS} = (\epsilon_f - \epsilon_{1s}) - E_b^F
    \label{eqn:simple_exciton}
\end{equation}
With this simplification, XAS peak energy differences result from:
(1) chemical shifts in the binding energy of the F 1$s$ orbital ($-\epsilon_{\rm 1s}$); 
(2) changes in the accessible (unoccupied) electronic band energies (within which $\epsilon_f$ resides), 
and (3) differences in exciton binding energy, as shown in Figure~\ref{fig:shifts}~(b).

The simplicity of this approximation has some drawbacks, particularly when studying condensed phases, where individual electronic states are buried in dispersive bands and electronic interactions can dramatically reorganize electronic structure due to strong mixing between orbitals  sampled from a continuum. However, our MBXAS calculations provide direct access to the various contributions of ground state orbitals to final state transitions via computed orbital overlap matrices between these two self-consisent fields. We can also estimate the exciton binding energy \textit{a posteriori} by mapping final state energies onto a linear combination of ground state electron-hole pairs (derived in the Supplementary Materials), as follows:
\begin{align}
    E_b^F = \frac{\sum_c |\bar{A}_{F,c}|^2 \epsilon_c}{\sum_c |\bar{A}_{F,c}|^2} - \tilde{\epsilon}_f \ ,
    \label{eq:Eb}
\end{align}
where $\epsilon_c$ are the ground-state orbital energies and $\bar{A}_{F,c}$ are the corresponding complex coefficients  given by $\braket{\Psi_{\rm{GS}}^{+c-\rm{core}}|\Psi_F}$, where $\ket{\Psi_F}$ is the many-electron final state while $\ket{\Psi_{\rm{GS}}^{+c-\rm{core}}}$ is obtained non-selfconsistently from the ground state by removing the core electron and adding it to the conduction level $c$.

In this manner, we can invert Eq.~\ref{eqn:simple_exciton} above to estimate the approximate origin of the XAS peak within the ground state density of states, based on our calculated binding energies:
\begin{equation}
    \epsilon_f = \epsilon_{1s}  + \Delta E_F^{\rm XAS} + E_b^F 
    \label{eq:ef_approx}
\end{equation}

This is illustrated in the middle panel of Fig.~\ref{fig:shifts}. Simply, the unoccupied electronic states contributing to the peak at $\Delta E_F^{\rm XAS}$ have an approximate energy $\epsilon_f$ located $\Delta E_F^{\rm XAS} + E_b^F$ above the absolute 1$s$ orbital energy. Examination of the ground state electronic density of states at this energy can reveal more about the nature of these electronic orbitals, as outlined below.

\paragraph{Core Electron Binding Energies} 
We use the differences in ground state Kohn-Sham (KS) potential at the F nuclei to capture the trend in electron binding energies. This is an initial state approximation, in which the effects of the core-hole and the associated electronic structure relaxation are not accounted for. To enable comparisons between different materials with different unit cells, we employ the same supercells as in our MBXAS calculations and reference the local KS potential at the F atom to that of an isolated F atom at the same position in the same supercell. With this common reference, comparisons between different materials or contexts should be valid (in particular for slab models of surfaces as outlined below).

In Fig. \ref{fig:shifts} we compare this local value of the potential (E$_{1s}$) with measured XPS binding energies~\cite{Moulder1992HandbookOX, wagner1991}. 
The calculated F 1$s$ relative  energies are aligned to a chosen experimental XPS binding energy reference. In this work, we chose that of F 1s in LiF~\cite{Moulder1992HandbookOX, wagner1991}. The negative binding energy is plotted in Fig.~\ref{fig:shifts}(b) to indicate the energetic position of the F 1s orbital with respect to the vacuum. We can see, for example, that it is $easier$ to extract a F 1$s$ core electron from NaF compared to LiF, due to the higher absolute energy (close to the vacuum level) or, equivalently, the lower binding energy of the former.

The XPS binding energy shows an approximate linear correlation with the local electrostatics around the excited atom, as has been noted for fluorides specifically.~\cite{kawamoto1999} The role of the electrostatic environment on the stabilization of the F 1$s$ core-hole can be estimated using Siegbahn's expression,~\cite{kawamoto1999} which describes the chemical shift $\Delta E_B$ of ion $i$ as a function of the charge distribution immediately surrounding it: 

\begin{equation}
    \Delta E_B(i) = k q_i + \sum_{j \neq i} \frac{e^2 q_j}{r_{ij}} + l
\label{eq:sieg}
\end{equation}

where $r_{ij}$  is the distance between charges $q_i$ and $q_j$. Here, we have used $r_{ij}$ obtained from the optimized crystal structures and Bader charges for $q_{i,j}$ (see Table~\ref{tab:sieg}). The trend in cation-anion distance follows that of the ionic radii of the cations.~\cite{crchandbook} The Bader charges on fluorine show little variation and are quite close to the nominal -1$e$ charge, with the exception of ZnF$_2$. In this species, hybridization in the ground state between the occupied Zn 3$d$ and F 2$p$ orbitals in the valence band indicates polarization of the fluoride anions by the Zn$^{2+}$ (at least within the PBE GGA calculations employed in this work).  We have fit Eq.~\ref{eq:sieg} to F 1$s$ XPS binding energies from the literature,~\cite{Moulder1992HandbookOX, wagner1991} obtaining $k$=3.99~eV e$^{-1}$ and $l$=685.9~eV; and reasonable agreement with the experimental XPS data (Table~\ref{tab:sieg} and Fig.~\ref{fig:shifts}).

As evident in Fig.~\ref{fig:shifts} (b), the calculated 1$s$ energies lie in a narrow range, close to -685~eV, with the experimental XPS energies displaying a slightly broader energy range, from -684 to -686~eV, followed somewhat by the Siegbahn model, with the most notable outliers being KF, MgF$_2$, and ZnF$_2$. However, this 1-2~eV range and the associated trends are insufficient to explain the observed $\sim$~4~eV range and non-monotonic trend in the XAS peak positions.

We note that we could have used the full core-hole (FCH) approximation to calculate electron binding energies, wherein the core-hole excited state is modeled using a modified pseudopotential~\cite{Pehlke1993} and relative excitation energies can be referenced within a given supercell or (as above) to an isolated atom. However, in these fluorides, we noticed significant binding energy underestimation, relative to LiF, in our FCH calculations for KF, MgF$_2$, ZnF$_2$ and, most prominently, CaF$_2$ compared to the XPS experimental binding energies and Siegbahn's electrostatic estimate. We might have expected this due to our choice of semi-local PBE exchange-correlation functional, which should overestimate electronic polarizability and underestimate the excitation energy. However, we recalculated the electron binding energies using hybrid functionals with different fractions of exact exchange and saw only small differences in the estimated relative binding energies with respect to PBE (see SI Section 2~\ref{sec:si_e1s}). Based on our surface studies (below) it is likely that the origin of this discrepancy is merely due to spurious electrostatic interactions between neighboring supercells, which converge slowly for these wide band gap insulators with little dielectric screening. We will explore solutions to this challenge in future studies.

\paragraph{Exciton Binding Energies}

The estimated exciton binding energies of the final states contributing most to the primary XAS peak of each fluoride are listed in Table~\ref{tab:sieg}. Despite the high energy of the excited state and the highly-localized F 1$s$ core orbitals, exciton binding energies are relatively small. The largest binding energies are for KF (0.99~eV) and CaF$_2$ (1.16~eV). In order to place exciton binding energies correctly on the energy scale, we added them to the excitation energy (obtained from the XCH method, see Computational methods).

\begin{table}[]
    \centering
    \begin{tabular}{cccccccccc}
          formula & q$_F$ & q$_M$ & r$_{FM}$ &  CN &   $\Delta E_B$  & XPS & $\Delta$XPS & $\Delta$E$_{1s}$ &  E$_b$ \\
         \hline
         LiF &-0.89 & 0.89 & 2.011&  6 & 685.0  & 685.0 & 0.00 & 0.00  & 0.15\\
         NaF &-0.85 & 0.85 & 2.352 & 6 & 684.7 & 684.5 & -0.5 & -0.04  & 0.50\\
         KF & -0.86 & 0.86 & 2.718 & 6 &   684.4& 683.9 & -1.1 & -0.15 & 0.99\\
        MgF$_2$ &-0.82 & 1.64 & 2.001 & 3 &685.1 & 685.7 & 0.7 & -0.37 & 0.73 \\
        CaF$_2$ &-0.88 & 1.76 & 2.387 &   4 & 685.3 &684.8& -0.2 & -0.23 & 1.16\\
         ZnF$_2$ &-0.74 & 1.48 & 2.066 &  3 & 685.1 & 685.1 & 0.1 & -0.49 & 0.42\\  
        \hline
    \end{tabular}
    
    \caption{Bader atomic charges, cation (M) - fluorine distance (in \AA), number of cations that the fluorine is coordinated to, and the resulting Siegbhan's binding energy ($\Delta E_B$), in reasonable agreement with the experimental XPS shift.\cite{Moulder1992HandbookOX}.  $\Delta$XPS provides the experimental relative binding energy with respect to LiF. $\Delta$E$_{1s}$ shows the change in electrostatic potential at the F nuclei relative to LiF. The last column indicates the calculated exciton binding energy for the primary XAS transition in each fluoride. All energies are in eV.}
    \label{tab:sieg}
\end{table}

Note that exciton binding energy, as defined in Eq.~\ref{eq:Eb} may not always be positive, even though we expect strong electrostatic binding between the core-hole and the excited electron. For example, orbital mixing in the final state may lead to contributions to the binding energy from both above and below the final state energy eigenvalue, which may even cancel to produce a binding energy close to zero or negative.
  
\begin{figure*}[ht]
\centering
\includegraphics[width=1.\linewidth]{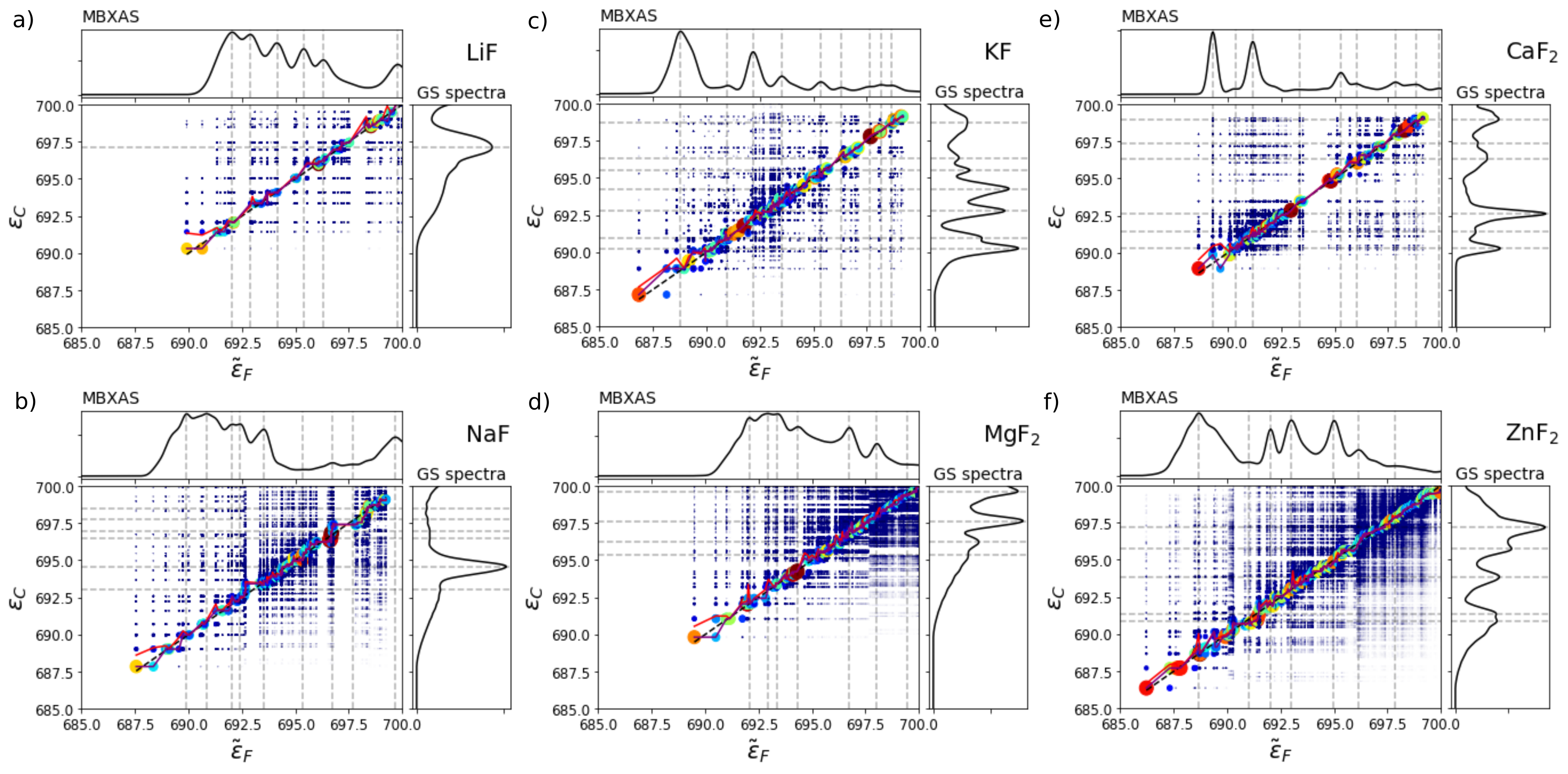}
\caption{ Values of $|\bar{A}_{F,c}|^2$ for the solids (a-f), plotted as a function of the final state energy ($\tilde{\epsilon}_F$) and the ground-state conduction energy ($\epsilon_C $), in eV.   Values of $|\bar{A}_{F,c}|^2$ larger than 10$^{-10}$ are shown both with color and point size. The black broken line marks $E_b^F$=0. The red line marks the total  $E_b^F$ (Eq. \ref{eq:Eb}), while the purple line marks a partial  $E_b^F$ calculated from only the largest $|\bar{A}_{F,c}|^2$ value for a given final  $\tilde{\epsilon}_F$.  On the top and right of each panel, the final spectra and the ground state spectra of each solid are shown.Gray dashed lines mark the maxima of the MBXAS and ground state spectra to guide the eye in connecting $\tilde{\epsilon}_F$ and  $\epsilon_C $ values. }
\label{fig:exciton_binding_map}
\end{figure*}

Within our example fluorides, we calculated a very small exciton binding energy for LiF compared to the other materials (0.15~eV). In this case the calculations are likely not fully converged: a larger supercell (4x4x4) gives E$_b$ = 0.8 eV. 
On the other hand,  due to the localized nature of the FCH unoccupied orbitals (and a different atomic structure), CaF$_2$ has a larger exciton binding energy.
A detailed plot of the single-particle states that contribute to the spectra and their corresponding exciton binding energies at the $\Gamma$ point (Fig.~\ref{fig:exciton_binding_map}) reveals the contrast between the LiF and the CaF$_2$ spectral first peak. While the first peak of CaF$_2$ is mainly contributed by one single-particle state with a very similar exciton binding energy, many states with negligible exciton binding energies contribute to the LiF first peak.

We would like to point out here that the peaks in the final state spectra cannot necessarily be simply mapped to peaks in the ground-state spectra (see Fig.~\ref{fig:siexciton}).
 An example is LiF, where the first main peak is contributed to by many final-state orbitals, each with a very small exciton binding energy (0.2~eV). This appears to be uncorrelated with the main peak in the ground-state spectrum, which is almost 5~eV higher in energy, as shown in Fig. \ref{fig:exciton_binding_map} (a). The largest values of $|\bar{A}_{F,c}|^2$ (see Eq. \ref{eq:Eb}) that contribute to the MBXAS first peak are due to initial-state orbitals originating in the pre-peak shoulder of the ground state spectrum. 
 
 In CaF$_2$, on the other hand, the states that contribute to the first peak of the final spectra have the largest  $|A_f|^2$ at the first peak of the ground state  spectra. Yet, the second peak (~ 691 eV) does not match with the first ground-state peak: instead, the largest  $|A_f|^2$ values correspond to the shoulder that sits between the ground-state main peaks. 
 
\begin{figure*}[ht]
\centering
\includegraphics[width=1.\linewidth]{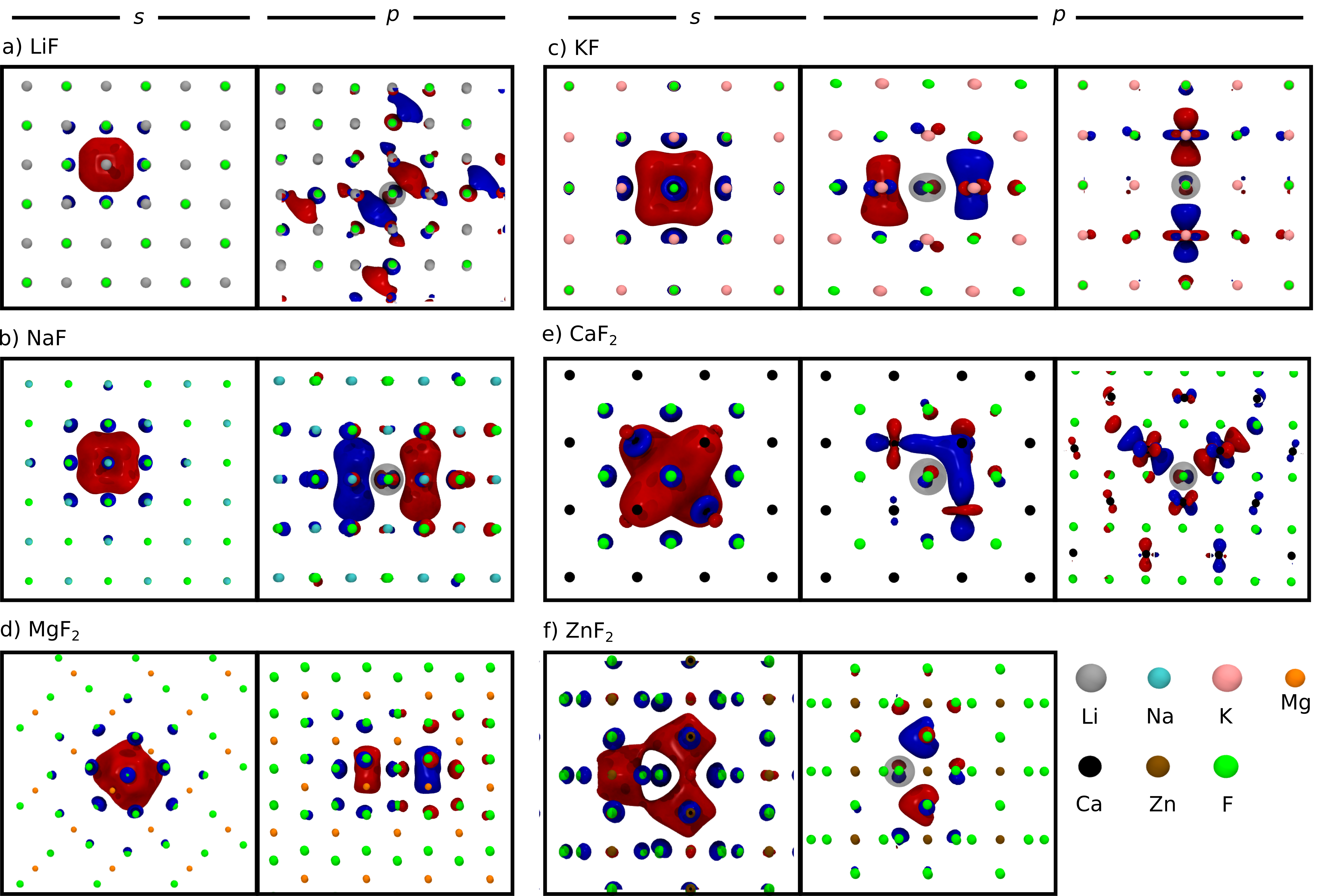}
\caption{Relevant single-particle states for the LiF, NaF, KF, MgF2, CaF$_2$ and ZnF$_2$ spectra: s-like state at the bottom of the valence band (a-f), main contributing state (F 3$p$) to peak (g-X). For each state, the isovalue is such that 30\% of the norm is contained within the isosurface. Shaded in gray is the core-excited atom FX.  }
\label{fig:sistates}
\end{figure*}

 Ultimately, the detail provided in Fig.~\ref{fig:exciton_binding_map} indicates how deficient the simple exciton model can be in describing the origins of X-ray spectral features in terms of ground-state orbitals and the care that should be taken before assigning the character of XAS peaks without checking their real-space representations (Fig.~\ref{fig:sistates}). 

\paragraph{Auxiliary ground-state unoccupied energy levels} Besides the shifts in the ground state F 1$s$ energy, the differences in final excitation energy can be traced  to the change in the ground-state orbitals that are involved in the formation of the exciton.  The simple exciton model uses an auxiliary energy level (red lines in Figure~\ref{fig:shifts} b)  to represent these, sitting E$_b$ above the final state energy, $\tilde{\epsilon}_F$. However, its energy need not coincide with the energy of any actual ground-state KS orbital. For instance, if the final state results from a combination of several ground-state orbitals with significant $|A_{F,c}|^2$ coefficients, all of them would contribute to the E$_b$ (Eq. \ref{eq:Eb}).

\subsection{Spectral Shape}

\bigskip

Due to the dipole-selection rule, the X-ray absorption spectra at the F $K$ edge (i.e., 1$s$ excitations) are defined by transitions to electronic orbitals with local $p$ ($l=1$) character at the core-excited fluorine (FX). In ionic solids such as these, the fluoride anion is nominally F$^-$ and the metal cations have their standard oxidation states (defined by the chemical formula): Li$^+$, Na$^+$, K$^+$, Ca$^{2+}$, Mg$^{2+}$, Zn$^{2+}$. To a first approximation, we would expect that the upper valence bands are predominantly F $2p$ in character, while the lower conduction bands have metal character, defined by their oxidation state ($s$ for all salts considered here). Therefore, one might expect XAS intensity to be limited to transitions to unoccupied orbitals with F $3p$ character at the core-excited F atom.~\cite{gao1993}  Integration of the FX $p$ projected density of states below the valence band maxima (VBM) produces values close to 6 (5.8 - 5.9) electrons, compatible with a mostly fully occupied 2$p$ band with little hybridization. 

The 3$p$ orbitals have been suggested, however, to be too high in energy, leading to low intensity transitions in these ionic solids.~\cite{vinogradov2005} In the core-excited state, there may be both downward shifts in the energy of orbitals with local FX $3p$ character and significant hybridization with available orbitals on nearby metal atoms (as well as other F atoms). And, if there is any significant hybridization between F and metal atoms in the occupied subspace, there may even be some F $2p$ character remaining in the low-lying unoccupied orbitals.

\begin{figure*}[ht]
\centering
\includegraphics[width=1.\linewidth]{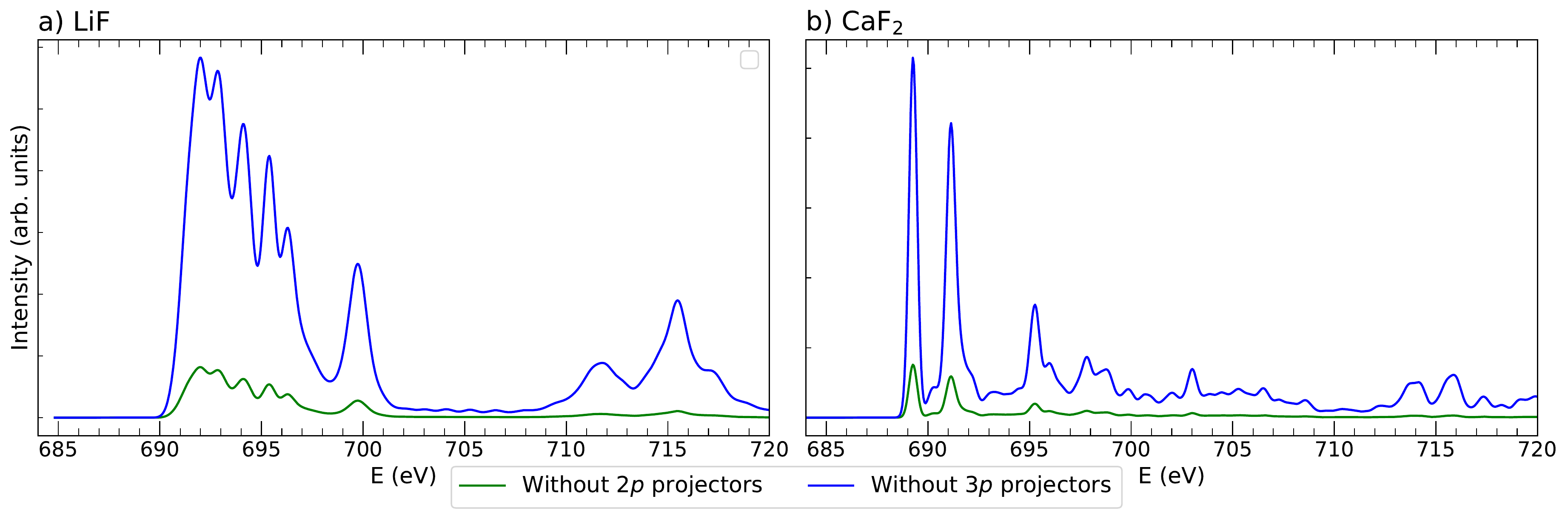}
\caption{Comparison of spectra of LiF (a) and CaF$_2$ (b) calculated without the contribution of 2$p$ or 3$p$ projectors, respectively.}
\label{fig:removed_projectors}
\end{figure*}

\paragraph*{pDOS analysis}

As stated above, the distribution of FX $p$ character in the conduction band is directly correlated with the XAS for a given crystal structure.  We would like to briefly comment here on the fact that including 3s and 3p projectors in the pseudopotential barely effects the lower near edge region of the spectra (Fig. \ref{fig:sippeffectspec}). However, a striking difference is observed on the fluorine atom pDOS. 
 The ground state fluorine pDOS near the conduction band minima generally maintains  F 2$p$ character (or mixed 2$p$ - 3$p$) while the core-excited state exhibits marked 3$s$ and 3$p$ character.  The presence of the core-hole stabilizes those 3$p$ states that lie at much higher energies in the ground state (see Fig. ~\ref{fig:sippeffect_pdos}). The negligible change of the spectra at lower energies, together with the striking differences in the ground-state and excited-state fluorine pDOS, suggest that the dominant transitions remain 1$s\rightarrow$2$p$ for all materials. Despite the majority of the fluorine pDOS being of 3$s$ and 3$p$ character in the lower energy region of the conduction band, the smaller spatial overlap of the 3$p$ states with the core-hole results in negligible oscillator strengths for these transitions. A direct assessment of the $p$ character of these transitions was carried out by eliminating alternatively the 2$p$ and 3$p$ projector contributions to the total spectra.  The results are shown in Fig. \ref{fig:removed_projectors} and indicate that most of the spectral intensity is due to 2$p$ character transitions, with a minor 3$p$ contribution. This is in partial agreement with the interpretation of Gao et al.\cite{gao1993} that the unoccupied states $and$ the spectral peaks of CaF$_2$ had 3$s$ and 3$p$ character. 

\begin{figure*}[htp]
\centering
\includegraphics[width=1.\linewidth]{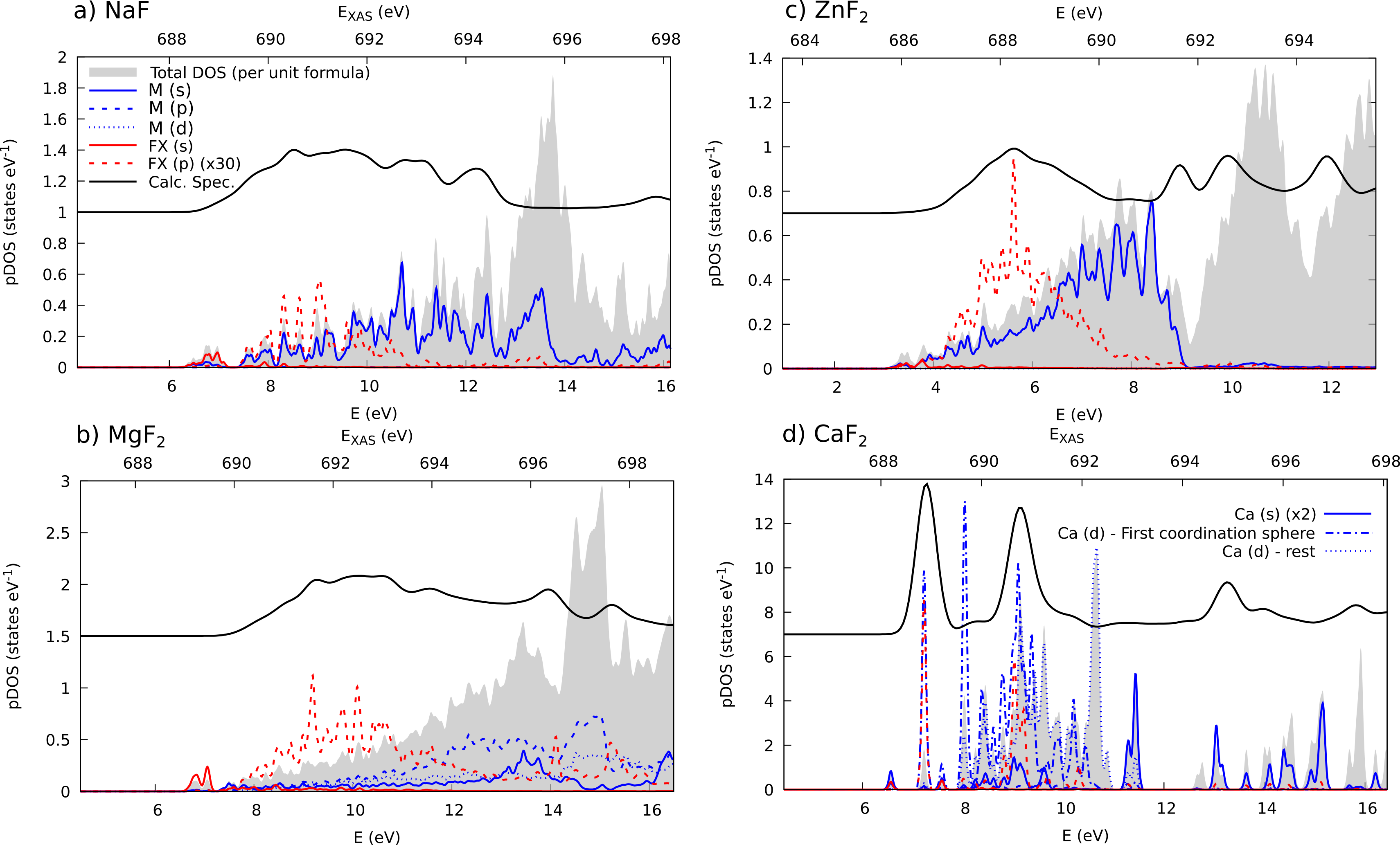}
\caption{Projected density of states in the conduction band of the core-excited state of NaF (a), MgF$_2$ (b), ZnF$_2$ (c) and CaF$_2$ (d). The total pDOS (per unit formula) is shaded gray, while the pDOS (per atom) of the cation and the core-excited F atom (FX) are shown in blue and red, respectively. The 2$p$ character of the FX pDOS is amplified 30 times. The pDOS energy scale is referenced to the valence band maxima of each material. To ease the comparison between the FX pDOS and the spectral lineshape, the calculated spectra of each material is shown as a black line, aligned to the FX pDOS peaks. }
\label{fig:condX}
\end{figure*}

Since the intensity of the near-edge is mostly due to excitations to states with 2$p$ character, Figure~\ref{fig:condX} shows the FX 2$p$ pDOS above the Fermi energy of NaF MgF$_2$,ZnF$_2$ and CaF$_2$, together with the metal pDOS.  Despite having different atomic structures (and, for ZnF$_2$, a fully occupied 3$d$-shell), all the considered fluorides have a wide FX 2$p$ band spanning close to 4~eV which coincides with the near-edge spectral features (red dashes in Fig.~\ref{fig:condX}) and overlaps with the metal $s$ character. As already observed for the spectral contributions in Fig.~\ref{fig:exciton_binding_map} and now for the pDOS in Fig.~\ref{fig:sicbm}, the FX $p$ core-excited states of LiF, NaF, MgF$_2$, and ZnF$_2$ sit above the conduction band minima -- i.e., these are resonant excitons.  

\bigskip
In CaF$_2$, spectral peaks stem from two narrow FX $p$ bands (Fig.~\ref{fig:condX}~(d)). In this case, integration of the FX $p$ pDOS below the VBM yields 5.8 electrons, which again indicates a mostly occupied 2$p$ band. However, mixing of FX and Ca orbitals is indicated by the strong difference in the 3$d$ pDOS of the FX first-neighbours compared to the remaining Ca atoms in the supercell (see blue dashed-dotted lines in Fig.~\ref{fig:condX}~(d)). Furthermore, comparing the $p$ pDOS of ground-state, core-excited and fluorine atoms far away from the core-excited atoms --that resemble the ground state-- shows a significant change in the energy of the FX $p$ pDOS peaks (Fig.~\ref{fig:si_pdos_distance}), in contrast to the other solids (with the exception of KF), again pointing to F $p$ - Ca 3$d$ mixing.  There is certainly 3 $d$ character in the valence band of CaF$_2$, amounting to 0.8 electrons, likely due to fluorine to metal mixing of F 2$p$ electrons with  unoccupied Ca 3$d$ electrons (see Fig. \ref{fig:si_pdos_valence_counter}). In comparison, integration of the nominally fully occupied 3$d$ orbitals of ZnF$_2$ returns 10 electrons. The spectra of KF are analogous to that of CaF$_2$, save for a few differences.  It should be noted that the bottom of the conduction band in KF does not show a significant 3$p$ character, but does show FX 3$s$, as observed in all the other solids (Fig.~\ref{fig:sicbm}). In CaF$_2$, the band-gap exciton and resonant exciton that lead to the double peak spectral feature have strong hybridization with the $e$ and $t_2$ 3$d$ orbitals of the surrounding Ca atoms, respectively (Fig.~\ref{fig:sistates}~(e)). Since the environment around the fluorine in KF is octahedral, instead of tetrahedral, the first and second peaks correspond to FX 3$p$ hybridization with K $t_{2g}$ and $e_g$ unoccupied 3$d$ orbitals (Fig.~\ref{fig:sistates}~(c)).

Note that in the analysis in Fig. \ref{fig:condX} the excited state energies are referenced to the top of the valence band, while those in Fig. \ref{fig:shifts} are referenced to the vacuum level.
The predominant $s$ character of the conduction band minima will be discussed below. 

\bigskip 

Visualization of key single-particle states can be misleading. Here, most have the appearance of 3$p$ states (Fig. \ref{fig:sistates}). However, as explained above, most of the spectral intensity arises from the mixed-in 2$p$ character.

\bigskip

In summary, despite the differences in atomic structure highlighted in Table~\ref{tab:latpars}, the two types of spectra -- main broad peak versus double-main peak -- can be attributed, respectively, to little hybridization and hence dominating F 2$p$ character (LiF, NaF, MgF$_2$, and ZnF$_2$) versus F 2\textit{p} - metal 3\textit{d } hybridization (KF, CaF$_2$).
Vinogradov \textit{et al.}~\cite{vinogradov2005} argued that for transition metal fluorides 3$p$ F states are too high above the ionization threshold and the origin of the peaks are transitions to lower-lying anti-bonding states arising from occupied F 2$p$ orbitals' hybridization with M 3$d$ orbitals. As 3d orbitals get occupied, the hybridized character gradually decreases and becomes negligible in ZnF$_2$.  While our findings regarding ZnF$_2$ are in agreement with Vinogradov \textit{et al.}'s, there is a clear contrast in the case of KF and CaF$_2$ (although the latter is not included in their study), where we observe 3$p$ - 3 $d$ hybridization in the conduction band. 

\bigskip

\paragraph*{Low-energy features at finite temperature}

Low-intensity peaks about $\sim$~2~eV below the main edge can be observed in most experimental XAS spectra in Fig.~\ref{fig:solid_spec}. In LiF~\cite{hamalainen2002,SCHWARTZ2017} and CaF$_2$,~\cite{gao1993} this weak feature has been assigned to dipole-forbidden $s$-to-$s$ transitions.  Computational studies have shown that, at finite temperature, local distortions modulate the octahedral symmetry about the fluorine atoms, turning a once `dark', dipole-forbidden $s$-to-$s$ transition into one with non-negligible oscillator strength in LiF.~\cite{SCHWARTZ2017} This effect has also been shown for the Li K edge of LiF.~\cite{pascal_2014_li}  In all fluorides studied here, and regardless of $p$ character, there is significant  F 3$s$-character  at the conduction band minima  (Fig.~\ref{fig:sicbm}).  Interestingly, the pervasive s-character at the conduction band minima indicates that these `dark' states are common to all ionic compounds studied here.  Finite-temperature spectra were obtained by averaging the spectra of each fluorine atom over multiple snapshots of an {\it ab-initio} molecular dynamics trajectory (see Computational Details), totalling above 200  individual atomic contributions to the spectrum of each material. Certainly, an increase in pre-peak intensity is observed in the finite-temperature averaged spectra (Fig.~\ref{fig:solid_spec}, green).
Additional features at even lower energies (e.g.  \~ 685~eV for LiF) have been assigned to radiation damage on the sample.~\citep{SCHWARTZ2017}

It should be noted that, besides the emergent pre-peak, temperature-averaged spectra and crystal structure spectra remain very similar for all structures studied here, with two exceptions. First, the differences between the crystal and temperature averaged spectra in LiF and NaF are due to the use of a smaller supercell (2x2x2) for the MD calculations, which, despite the additional exciton localization provided by local distortions in the F environment, may still be too small.  Secondly, the first-to-second peak ratio in CaF$_2$ temperature-averaged spectra differs from the crystal structure spectra. The "true" exciton responsible for the first peak in CaF$_2$ is slightly more susceptible to finite temperature effects than the resonance exciton that leads to the second peak, resulting in a moderate decrease in intensity of the former compared to the latter.

\bigskip

\paragraph*{Effects of DFT functional}The choice of exchange-correlation functional affects the spectral shape too. Local and Semi-local functionals, such as PBE used here, notoriously suffer from self-interaction error. This leads to band gap and band width underestimation,~\cite{mori2008} in particular in species with low dielectric screening. LiF has the widest band gap in comparison with the remaining solids, and hence is the most affected: the spacing between peaks in the LiF spectra is too narrow. Agreement with experiment greatly improves by uniformly stretching the spectra by a factor of 1.2. 
Additionally, $d$ orbitals are particularly affected by self-interaction error (as compared to $s$ and $p$ orbitals). The use of a semi-local functional such as PBE may cause $d$ orbital over-delocalization, which can explain the fact that the second peak in KF and CaF$_2$ calculated spectra appear at a lower energy than experiment. 

\bigskip

\paragraph*{Many-body effects}The consequences of including many-body effects in evaluation of the transition amplitudes are non-trivial (see Fig.~\ref{fig:sicalcall}) . An increase in intensity is generally observed at the main edge and lower energy peaks with respect to the higher energy peaks. This is due to non-zero off-diagonal elements in the overlap matrix between the initial and final state orbitals, which lead to non-zero intensity for transitions that would be forbidden in the single-particle picture.~\cite{Roychoudhury2021} Including many-body effects in this manner generally leads to better agreement with experimental intensities. 

The inclusion of many-body effects increases the lower energy peak intensities for both rutile structures, MgF$_2$ and ZnF$_2$, when comparing with single-particle calculations. This increase in low energy intensity is in better agreement with experiment. The underestimated low energy intensity was already seen in similar single-particle calculations provided by Yamamoto {\it et al.}\cite{Yamamoto2004}

We note that for CaF$_2$, the relative intensity between the first two main peaks seems to be dependent on the experimental conditions. The progressive introduction of electronic structure nuances in the calculation leads to an increase in lower-energy peak intensity, in better agreement with our experimental spectra. That is, taking into account the core-hole, many-body effects (as in MBXAS) and finally finite temperature effects steadily improves the computational results (see Fig.~\ref{fig:sicalcall})

\subsection{Surface contributions to XAS}

We explored the variation of the F K-edge spectra on exposed surfaces, where the fluorine atoms can be under-coordinated and immersed in a different local electrostatic environment.  

\begin{figure*}[htp]
\centering
\includegraphics[width=0.9\linewidth]{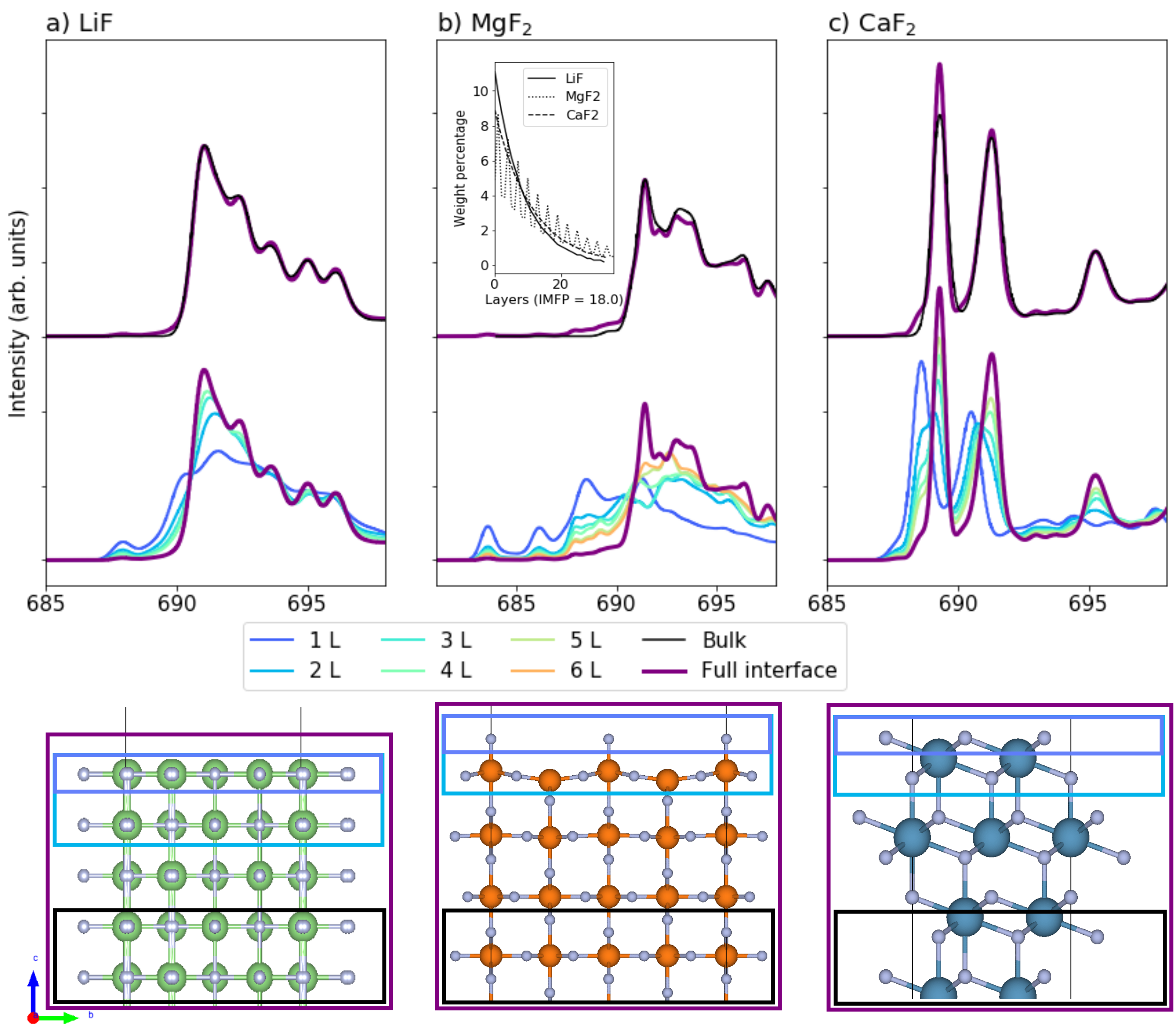}
\caption{Side view of the atomic structures of the model slabs: LiF(001) (left), MgF$_2$(110) (center) and CaF$_2$ (111) (right), with some example integration regions highlighted in mauve (first layer), blue (first and second layer), black (bulk) and purple (full interface), respectively. Above, this color scheme is used to show the integrated spectra of the corresponding atomic layers. The bulk and estimated interfacial spectra (full interface) are shown at the top in black and purple.  }
\label{fig:surf}
\end{figure*}

We compare here the most stable low index surfaces of cubic, rutile, and fluorite solids: LiF(001), MgF$_2$(110)~\cite{Huesgues2013,Han2017,WangY2020} and the hydrophobic CaF$_2$(111), surface with Type 2 termination (see Computational Details), which was determined to be the most stable under dry conditions.~\cite{ZhangWangMiller2015} While the cubic (001) surface is apolar, the rutile (110) and fluorite (111) surfaces have a surface dipole (see Computational Details). The atomic structural details are shown in Figure~\ref{fig:surf}. Surface relaxation leads to minimal distortion in LiF and CaF$_2$ (a slight elongation of Ca-F distance in the topmost layer by about 0.05 \AA). On the other hand, in  MgF$_2$, the Mg atoms in the outermost layers distort out-of-plane, with those that are under-coordinated displaced towards the center of the slab.

\paragraph*{Interfacial Spectra} The first step to obtain interfacial atomic spectra was to calculate the spectrum of each individual atomic layer for each surface. We note that in CaF$_2$ (MgF$_2$),  the third (seventh) layer spectrum is converged to that of the bulk. 
The difference between the first layer spectrum (1L, mauve in Fig.~\ref{fig:surf}) and the bulk spectrum (black, top in Fig.~\ref{fig:surf} ) in all surfaces is striking.

The new spectral features from top surface layers that appear well below the absorption edge are not a result of electrostatic perturbations. The electrostatic potential of the ground state and with a core-excited atom reaches a flat plateau in the vacuum level in all cases (see Fig.~\ref{fig:sislab_pot}). 
Lower intensity results from a slightly larger delocalization of those F 2$p$ contributing orbitals that are perpendicular to the surface. Their significant red-shift may also result from under-coordination, particularly in the case of MgF$_2$(110) (Fig.~\ref{fig:surf}, blue). 

In a more general context, these results indicate that lower energy peaks (relative to spectral standards for ionic solids) can be due the presence of interfacial F atoms. However, to what extent these peaks would be noticeable in an experimental setup requires clarification. Taking into account the mean free path (IMFP) of electrons at this energy range (close to 1.8 nm),~\cite{tanuma1991, Flores-Mancera2020, boutboul1996} we have generated a series of estimates (see Computational Methods) of the interfacial spectra of these surfaces that might be produced using a surface-sensitive XAS measurement such as TEY (Figure~\ref{fig:surf}). We show spectra that steadily include contributions from greater depth (with exponential decay in these contributions consistent with the IMFP). Spectra where contributions from two to six layers are detected show a gradual change where the pre-edge intensity decreases. Although the red-shifted peaks are most prominent up to second-layer detection, they are still noticeable when the full interface is taken into account, especially in MgF$_2$. 

Our MgF$_2$ results can explain differences in measured spectra between the bulk-sensitive transmission experiments of Oizumi {\it et al.},~\cite{oizumi1985} where very little pre-edge intensity is detected (perhaps only that related to temperature-enabled transitions); and the surface sensitive TEY experiments of Yamamoto {\it et al.},~\cite{Yamamoto2004} where a broad-pre-edge is observed (Fig.~\ref{fig:solid_spec}). 

\begin{figure*}[htp]
\centering
\includegraphics[width=0.9\linewidth]{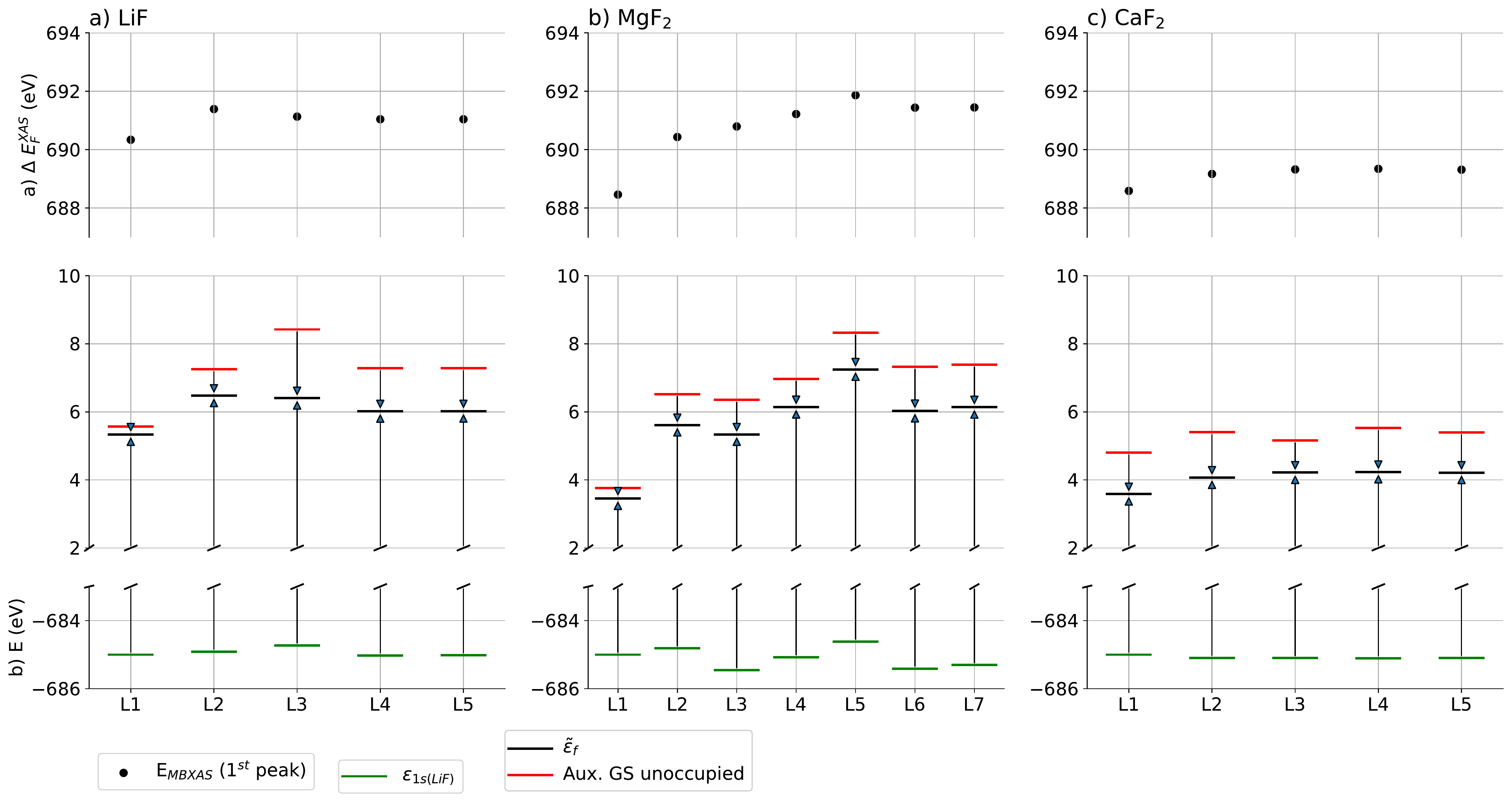}
\caption{Scheme showing  the changes in   $\epsilon_{1s}$, $\epsilon_f$, $E_b^F$, and  $\Delta E_F^{\rm XAS}$    (Eq. \ref{eq:ef_approx})  that contribute to the XAS relative alignment for each layer of a slab of   LiF(001)  (a), MgF$_2$ (110) (b) and CaF$_2$ (111) (c), with L1 being the outermost layer and L5 (L7) the center of the slab. }
\label{fig:surf_diagram}
\end{figure*}

Ion-yield measurements of  CaF$_2$(111) by Tanaka et al.,~\cite{Tanaka_caf2_surf_1998} sensitive only to the first layer of the surface, indicate a completely different lineshape, with enhanced first peak intensity and an overall estimated red shift of $\sim$~-1.2 eV and , in pretty good agreement with our own results for the top-most layer of the same surface (-0.7 eV red shift and ehancement of the first peak).
Changes in the surface Madelung potential were considered the main source of the red-shift~\cite{Tanaka_caf2_surf_1998}, but the local potential at the excited atom hardly changes (Fig.~\ref{fig:surf_diagram}), while the X-ray excited state energy drops considerably.

The recognizable sharp double-peak feature in the spectrum of CaF$_2$ also sees further changes in intensity ratio due to surface effects. We had already noticed that this intensity ratio favored an increase in intensity of the first peak due to inclusion of many-body contributions to the transition intensity and the influence of finite temperature molecular dynamics sampling. Now, we also notice an enhancement of the first peak due to surviving surface contributions to the overall surface-sensitive (TEY) spectrum, which is not the case in the measurement of Oizumi et al.~\cite{oizumi1985}. 

The sensitivity of our estimated  interfacial spectra to the value of the IMFP is shown in Fig.~\ref{fig:siinterface_imfp}. Interestingly, lower energy features do not completely disappear even by increasing the IMFP to 3 nm. On the other hand, a dramatic increase in their intensity can be observed when the IMFP is reduced. We propose that our calculations can serve as guidance in the estimation of material-specific IMFP values from experiment, by specific analysis of pre-edge intensity. 

Overall, low-intensity features below the main edge together with  a slight change in peak intensity ratio (in the case of CaF$_2$) seem to be the key signatures of interfacial F atoms. More generally, lower energy features can be expected when two conditions are met: a short enough IMFP (e.g., below 4 nm), and a significantly different surface layer(s) spectrum due to uncoordinated atoms. 

\paragraph*{Exciton binding at the surface} We have used the strategy in Section A to produce an alignment diagram for each surface that describes resonant excitations of core-excited atoms from the center of the slab towards the surface, shown in Figure \ref{fig:surf_diagram}. 
The effect of the local electrostatic environment (reflecting the F 1$s$ binding energy) is small, especially for LiF and CaF$_2$, but noticeable for MgF$_2$, due to the inequivalent positions of the F layers with respect to the surface dipole and under-coordinated Mg atoms.

Exciton binding energy analysis indicates that the exciton binding energies of LiF are not converged with respect to supercell size, even at these large sizes. For the bulk, $E_b$ increases from 0.3 to 0.8 eV from 3x3x3 to 4x4x4; with ground state orbitals that contribute to the final state more distributed in energy in the latter case. Furthermore, comparing the center of 3x3x4 and 3x3x5 slabs indicates vanishing contributions to the oscillator strength for polarizations perpendicular to the surface, and larger binding energies (close to 2 eV), with, again, a wider distribution in the $|A_{F,c}|^2$ intensities in the thicker slab.

Given the large computational expense of these supercell calculations, methodological improvements may be necessary to remove these convergence issues for LiF (and other weakly screened ionic solids).

\paragraph*{Future surface-sensitive experiments} Even though we have shown the existence of well-defined pre-edge spectral features that originate only from surface atoms, these features are not as prominent in X-ray absorption due to the contributions of more atoms from bulk-like environments below the surface accessible due to the IMFP of emergent electrons. However, the existence and spectral isolation below the main-edge of these surface states admits the possibility of enhancing the surface or interfacial sensitivity of nominally bulk sensitive measurements. For example, resonant experiments, such as Resonant X-ray Emission Spectroscopy (RXES) or Resonant Inelastic X-ray Scattering (RIXS) could be utilized with incident X-rays tuned to these energy ranges where surface states exist -- well below the absorption edge. With sufficient detector sensitivity, this should reveal emission signals indicative of the local electronic structure about these under-coordinated or otherwise-modified atoms. Additionally, spectra with high spatial resolution can be measured using electron energy loss spectroscopy (EELS). If samples could be oriented such that the electron beam runs along an exposed salt surface, then such surface contributions could be analyzed with almost atomic resolution. Granted such experiments may be quite challenging due to beam-induced damage, sample charging, etc.

\subsection{Conclusions}

Despite the fact that the oxidation state of fluorine remains mostly unchanged from one fluoride salt to another (as the most electronegative element) we find richness in its excited-state electronic structure in different materials as evident through F K-edge X-ray absorption spectra (XAS). First-principles analysis using density functional theory using a many-body formalism developed previously for XAS prediction (so-called MBXAS)~\cite{liang2018,liang2019} indicates and explains significant differences in XAS of the fluoride salts: LiF, NaF, KF, MgF$_2$, CaF$_2$, and ZnF$_2$. Our theoretical work highlights inconsistency in the experimental literature with respect to spectral alignment, which is a useful caveat for future measurements at the F K edge of less well-defined samples. Consistent alignment is vital for interpretation using a predictive theory, particularly for systems that may exhibit mixed fluorine chemistry - both ordered and disordered inorganic fluorides, fluorinated organics, etc. Despite this, the spectral comparisons between theory and experiment are excellent and encourage more detailed studies of the physical origins of subtle differences between theory and experiment that might relate to the measurement or sample details, or between the spectra of distinct systems and their inherent atomic and electronic properties. Specifically, here we focused on observed and predicted shifts in the first/main XAS peak energies in terms of a simple exciton model. This approach highlighted the importance of excited-state orbital mixing with metal character, in particular the (empty) $d$ character of Ca and K. Furthermore, a detailed understanding of distinct spectral signatures of materials surfaces is provided, indicating that strongly red-shifted peaks result from reduced coordination and electrostatics at salt surfaces -- again, highlighting the importance of well-calibrated measurements to capture such effects in interfacially sensitive measurements. Overall, the analysis presented here indicates that the F K edge provides rich information on fluorine chemistry that can complement studies of evolving materials chemistry in a number of relevant geochemical and technological contexts. 

\section*{Acknowledgments}

This work was supported primarily by the Joint Center for Energy Storage Research, an Energy Innovation Hub funded by the United States Department of Energy, Office of Science, and Basic Energy Sciences. The theoretical analysis in this work was supported by a User Project at The Molecular Foundry and its computing resources, managed by the High Performance Computing Services Group at Lawrence Berkeley National Laboratory (LBNL), and the computing resources of the National Energy Research Scientific Computing Center, LBNL. XAS measurements were made using resources of the Advanced Light Source, a DOE Office of Science User facility at LBNL. All LBNL facilities employed in this work were supported by the Director, Office of Science, Office of Basic Energy Sciences, of the United States Department of Energy under Contract DE-AC02-05CH11231.

Sandia National Laboratories is a multimission laboratory managed and operated by National Technology \& Engineering Solutions of Sandia, LLC, a wholly
owned subsidiary of Honeywell International Inc., for the U.S. Department of Energy’s National Nuclear Security Administration under Contract DE-NA0003525. This paper describes objective technical results and analysis. Any subjective views or opinions that might be expressed in the paper do not necessarily represent the views of the U.S. Department of Energy or the United States Government.

\bigskip

\textbf{Supporting Information Available: }Derivation of the exciton binding energy, as well as plots showing the effect on spectra of calculation variables (supercell size, type of pseudopotential, level of theory,...). 
This material is available free of charge via the Internet at
http://pubs.acs.org

\bibliography{main}%

\end{document}


\setcounter{equation}{0}
\setcounter{figure}{0}
\setcounter{table}{0}
\setcounter{page}{1}
\makeatletter
\setcounter{section}{0}

\renewcommand{\theequation}{S\arabic{equation}}
\renewcommand{\thefigure}{S\arabic{figure}}
\renewcommand{\thetable}{S\arabic{table}}
\renewcommand{\bibnumfmt}[1]{[S#1]}

\section{Section 1: Derivation of the exciton binding energy }

As outlined in the main manuscript, in order to make reliable assignments of our X-ray spectral features to corresponding ground state electonic structure, we initially employ a simple exciton model that defines three key contributions to the X-ray excitation energy. One of these is the core exciton binding energy, which is one measure of how much the ground state electronic structure is reorganized due to the core-hole. We capture such reorganization immediately within our final state core-hole calculations that define the final state orbital energies, but we lack the direct connection to the initial state electronic structure. Fortunately, the MBXAS formalism expresses each core-excited state orbital as a linear combination of initial state orbitals, via the orbital overlaps. There is a direct analogy between this approach and the Tamm-Dancoff approximation for excited states within the Bethe-Salpeter equation (Ref. 20 of the main manuscript). We can use this analogy to derive an expression for the binding energy that is readily calculable with quantities available from our MBXAS method.

The Bethe-Salpeter equation (BSE) relies on the construction and diagonalization of the two-particle Hamiltonian $\hat{H}^{2p}$ whose $F$-th eigenvalue $\Omega^F$ is the $F$-th excitation energy and the corresponding eigenvector
\begin{align}
    \ket{\Psi_F} = \sum_{v,c} A^F_{v,c}\ket{v}\otimes\ket{c}
    \label{BSEEigVec}
\end{align}
can be expressed as a linear-combination of pairs of a valence ($v$) hole and a conduction ($c$) electron taken from the reference (ground) state. The excitation energy can then be written as
\begin{align*}
    \Omega^F &{}= \frac{\braket{\Psi_F|\hat{H}^{2p}|\Psi_F}}{\braket{\Psi_F|\Psi_F}}\\ &{} = \frac{\sum_{v,c}(\epsilon_c - \epsilon_v)|A^F_{v,c}|^2 + \sum_{v,c,c',v'}K_{v,c,c',v'} A^F_{v,c} A^{F*}_{c',v'}}{\braket{\Psi_F|\Psi_F}}, 
\end{align*}
where $\epsilon$ denotes quasiparticle (QP) energy and the BSE kernel $\hat{K}$ incorporates the explicit electron-hole (e-h) interaction. 
Roughly speaking, the binding energy ($E_b$) is obtained by subtracting the excitation energy from the QP energy difference, which, as expected, does not include any e-h attraction. 
However, since each excitation is composed of multiple states each of which contains a different e-h pair (see Eq. \ref{BSEEigVec}), a one-to-one map can not be drawn between
$\Omega^F$ and the QP difference. 
Therefore, in the spirit of the BSE, the following expression can be assigned to $E_b$:
\begin{align}
    E_b^F = \frac{\sum_{v,c}(\epsilon_c - \epsilon_v)|A^F_{v,c}|^2}{\sum_{v,c} |A^F_{v,c}|^2} - \Omega^F
    \label{BSE_Eb}
\end{align}
In our calculation, instead of the BSE, we use the inexpensive MBXAS formalism where the QP orbitals are replaced by the Kohn-Sham (KS) counterparts and for the $F$-th excited state (say), all the KS orbitals are obtained self-consistently in the presence of a core-hole such that, in addition to the bottom-most $N$ valence orbitals, the $f$-th ($f > N$) conduction orbital with energy $\tilde{\epsilon}_f$ is occupied, where $N$ is the number of valence electrons in the ground state.

\begin{align}
    \ket{\Psi_F} = \sum_c \braket{\Psi_{\rm{GS}}^{+c-\rm{core}}|\Psi_F}\ket{\Psi_{\rm{GS}}^{+c-\rm{core}}} = \sum_c \bar{A}_{F,c} \ket{\Psi_{\rm{GS}}^{+c-\rm{core}}},
\end{align}
where the sum runs over all the unoccupied levels of the ground state and
\begin{align}
    \ket{\Psi_{\rm{GS}}^{+c-\rm{core}}} = \hat{a}^\dagger_c \hat{a}_{\rm{core}} \ket{\Psi_{\rm{GS}}},
\end{align}
such that $\ket{\Psi_{\rm{GS}}}$ is the ground-state and $\hat{a}$ stands for the electron annihilation operator. Then, in Eq.\ref{BSE_Eb}, making the substitutions $A^F_{v,c} \rightarrow \bar{A}_{F,c}$ , $\epsilon_v \rightarrow \epsilon_{\rm{core}}$ , $\Omega^F \rightarrow (\tilde{\epsilon}_f - \epsilon_{\rm{core}})$, we obtain
\begin{align}
    E_b^F = \frac{\sum_c |\bar{A}_{F,c}|^2 \epsilon_c}{\sum_c |\bar{A}_{F,c}|^2} - \tilde{\epsilon}_f
\end{align}

\begin{figure*}[ht]
\centering
\includegraphics[width=1.\linewidth]{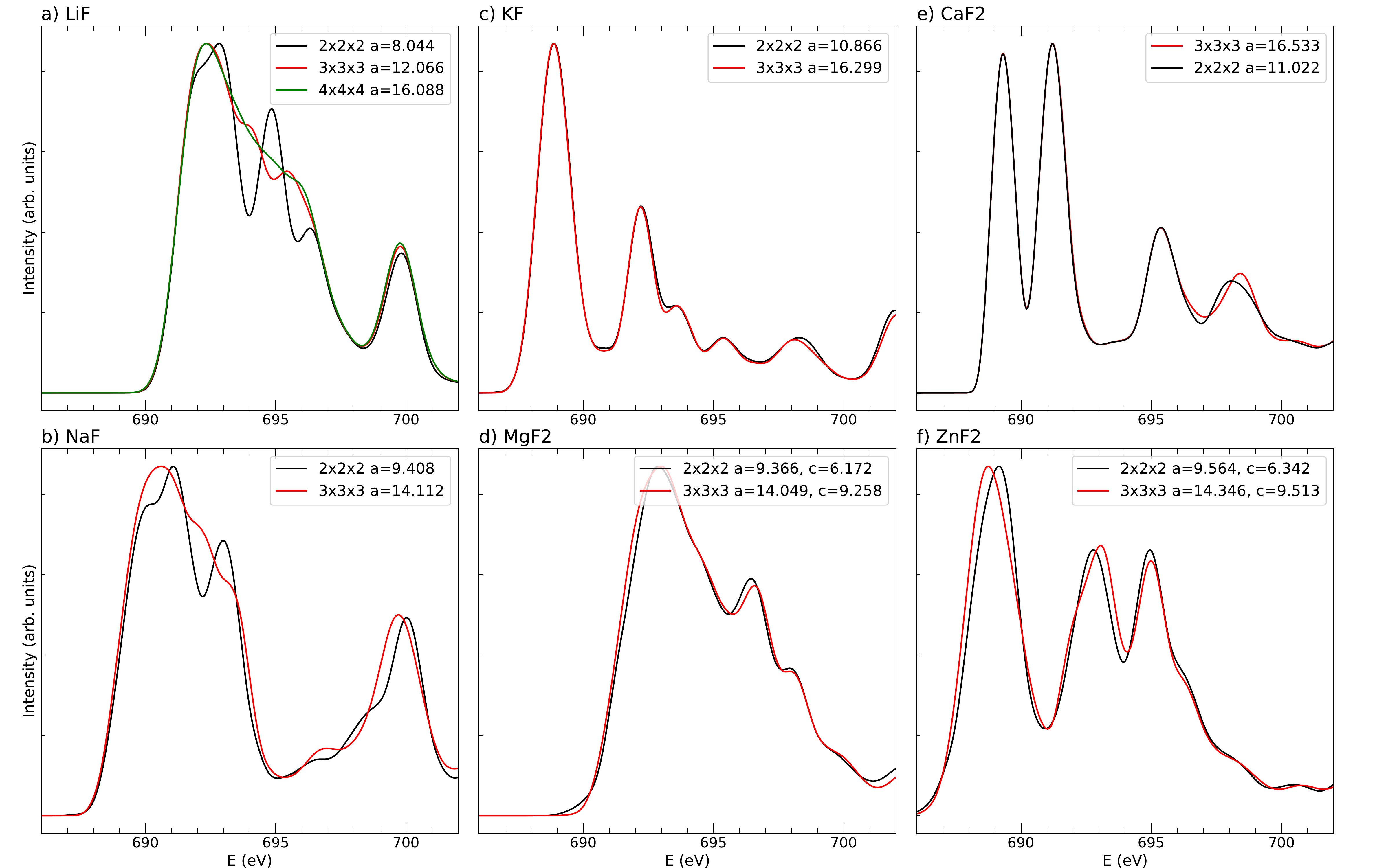}
\caption{Calculated spectra of reference solids using 2x2x2 (black),  3x3x3 (red) and, for LiF , 4x4x4 supercells, with the corresponding lattice parameters (\AA).}
\label{fig:si_super}
\end{figure*}

\begin{figure*}[ht]
\centering
\includegraphics[width=1.\linewidth]{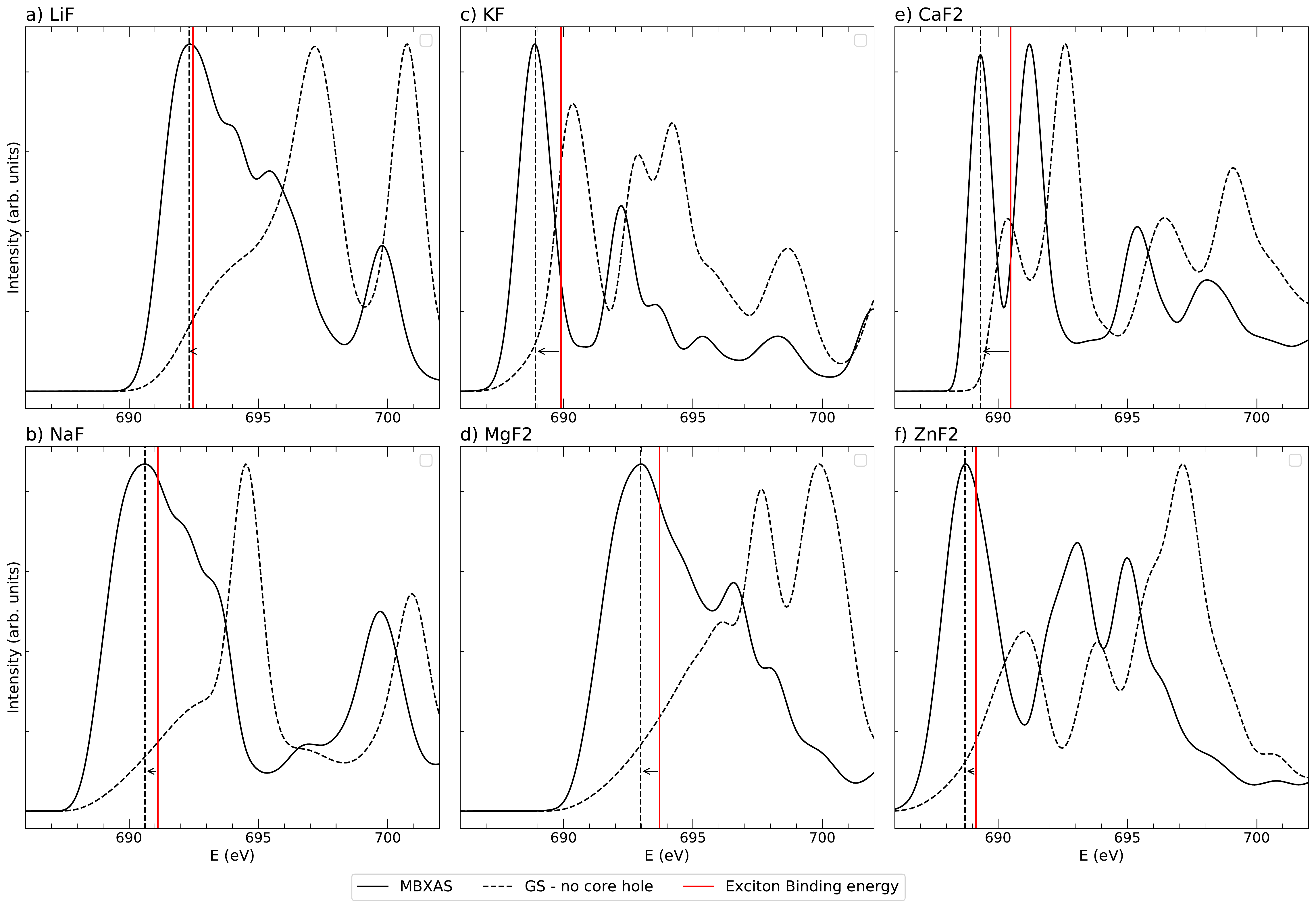}
\caption{Comparison of the ground-state and MBXAS spectra spectra of all solids, in dashed and black lines, respectively. The main MBXAS peak is highlighted with a black line. The exciton binding energy the main contributing state to that peak is added to that, resulting on the red vertical line. While the match between the ground-state main peak and the MBXAS main peak, mediated by the exciton binding energy is in great agreement in CaF$_2$, that is not the case for the remaining materials. }
\label{fig:siexciton}
\end{figure*}

\begin{figure*}[ht]
\centering
\includegraphics[width=1.\linewidth]{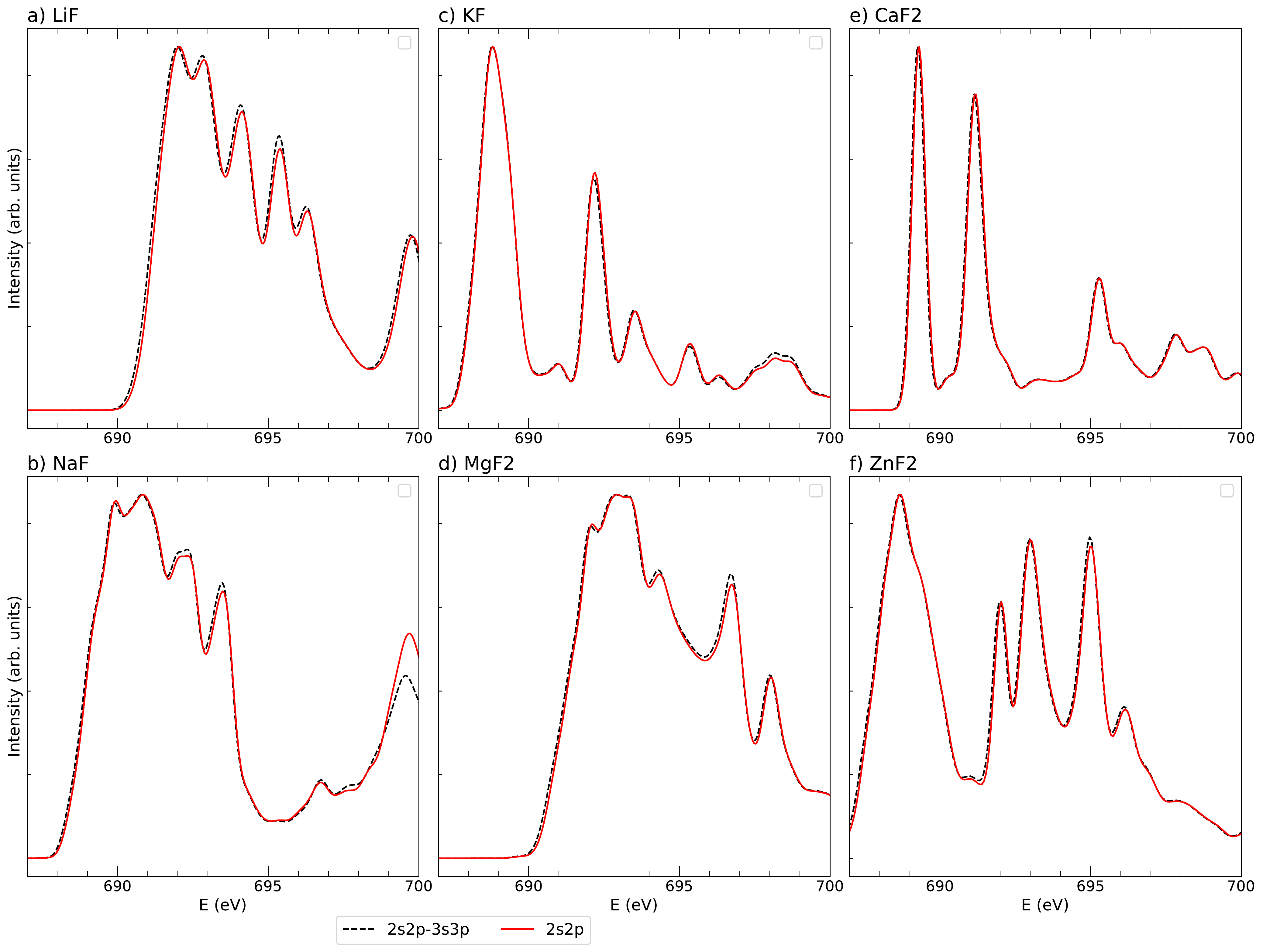}
\caption{Effect on the calculated spectra of including 3s and 3p projectors in the Fluorine pseudopotentials.}
\label{fig:sippeffectspec}
\end{figure*}

\begin{figure*}[ht]
\centering
\includegraphics[width=1.\linewidth]{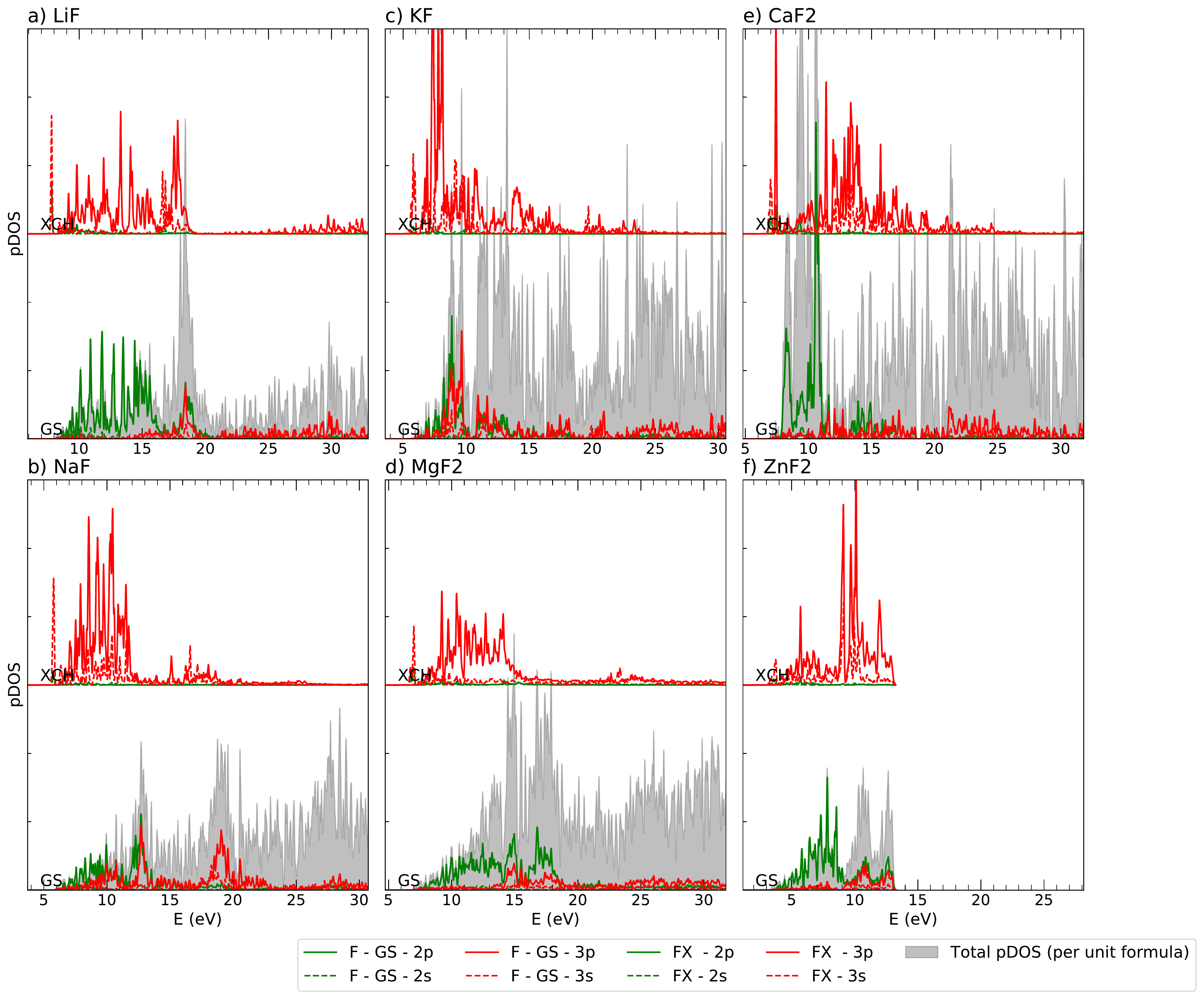}
\caption{Projected density of states for ground state (bottom) and core-excited F (FX, top) in LiF, NaF, KF, MgF$_2$, CaF$_2$ and ZnF$_2$ separated into 2$p$ and 3$p$ contributions; this highlights the stabilization of the 3$p$ states in the core-excited fluorine atoms.    }
\label{fig:sippeffect_pdos}
\end{figure*}

\begin{figure*}[ht]

\includegraphics[width=1.\linewidth]{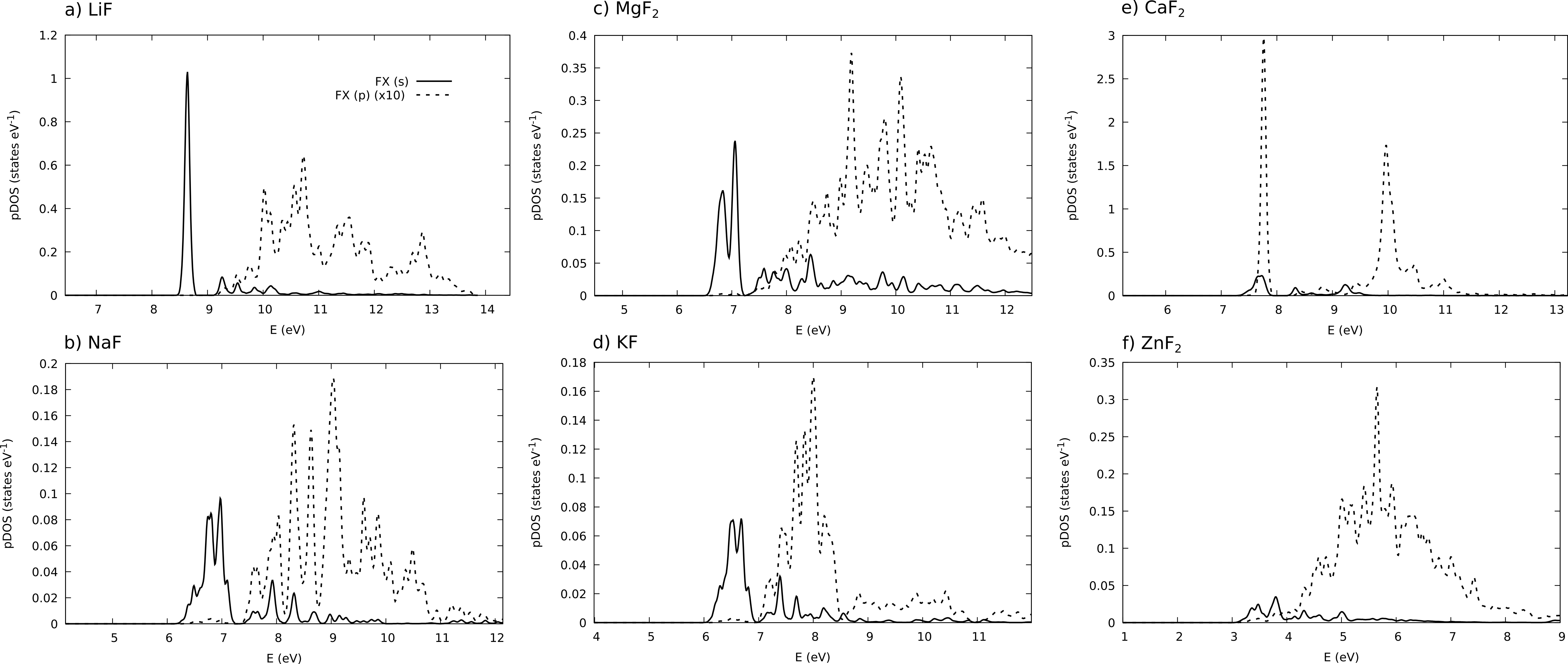}
\caption{Bottom of the conduction band of LiF, NaF, KF, MgF$_2$, CaF$_2$ and ZnF2. }
\label{fig:sicbm}
\end{figure*}

\begin{figure*}[ht]
\centering
\includegraphics[width=1.\linewidth]{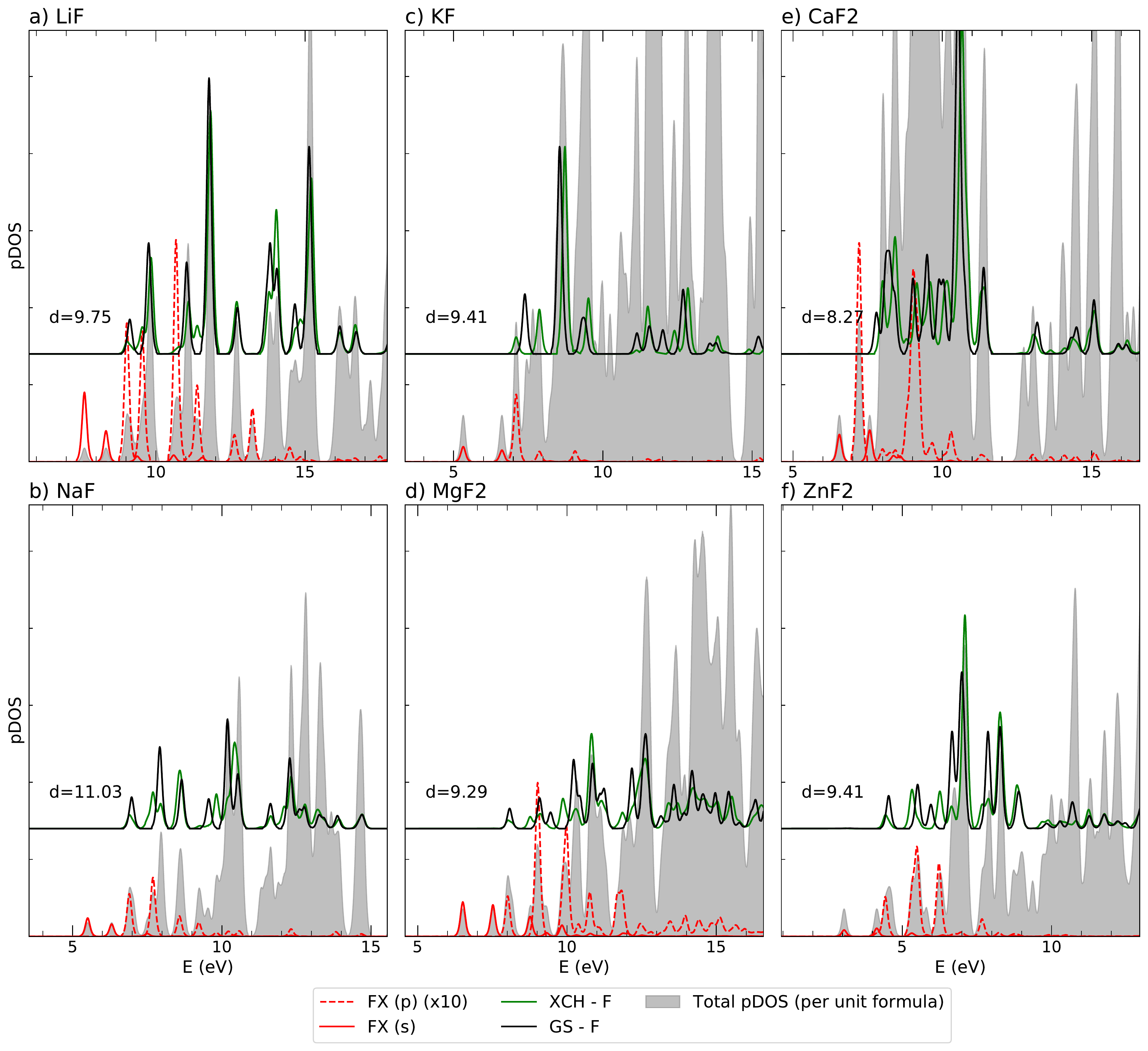}
\caption{Comparison of the $p$ projected density of states (pDOS) of the core-excited fluorine (red) with other (non-excited) fluorine atoms (green) as a function of distance (in angstrom) from the FX. The ground state fluorine pDOS is also given for comparison (black).   }
\label{fig:si_pdos_distance}
\end{figure*}

\begin{figure*}[ht]
\centering
\includegraphics[width=1.\linewidth]{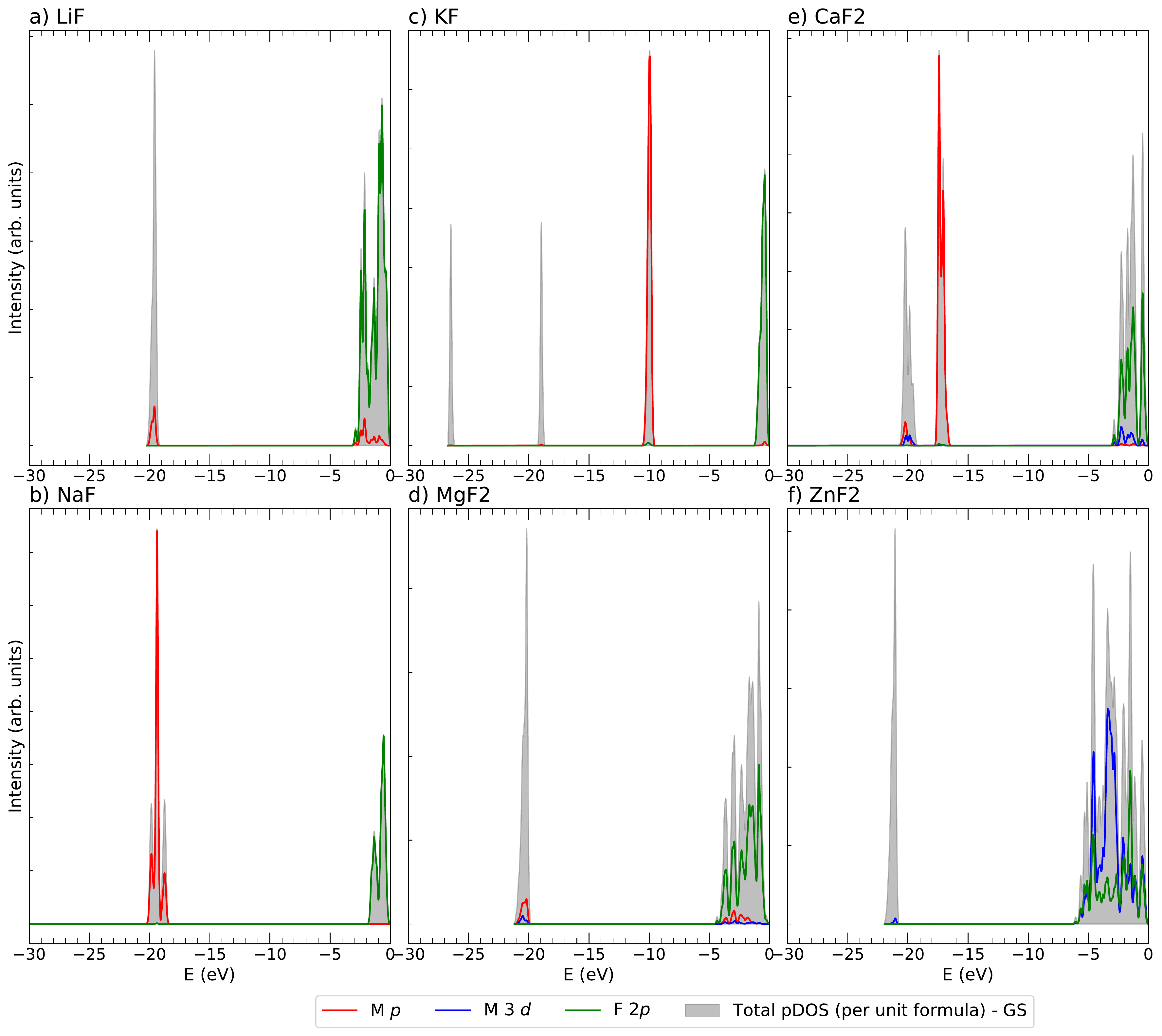}
\caption{Comparison of the projected density of states (pDOS) below the Fermi level with F 2$p$ character (green), metal p character (red), and metal $d$ character (blue) the core-excited fluorine (red) with other (non-excited) fluorine atoms (green) as a function of distance (in angstrom) from the FX. The ground state fluorine pDOS is also given for comparison (black).   }
\label{fig:si_pdos_valence_counter}
\end{figure*}

\begin{figure*}[ht]
\centering
\includegraphics[width=1.\linewidth]{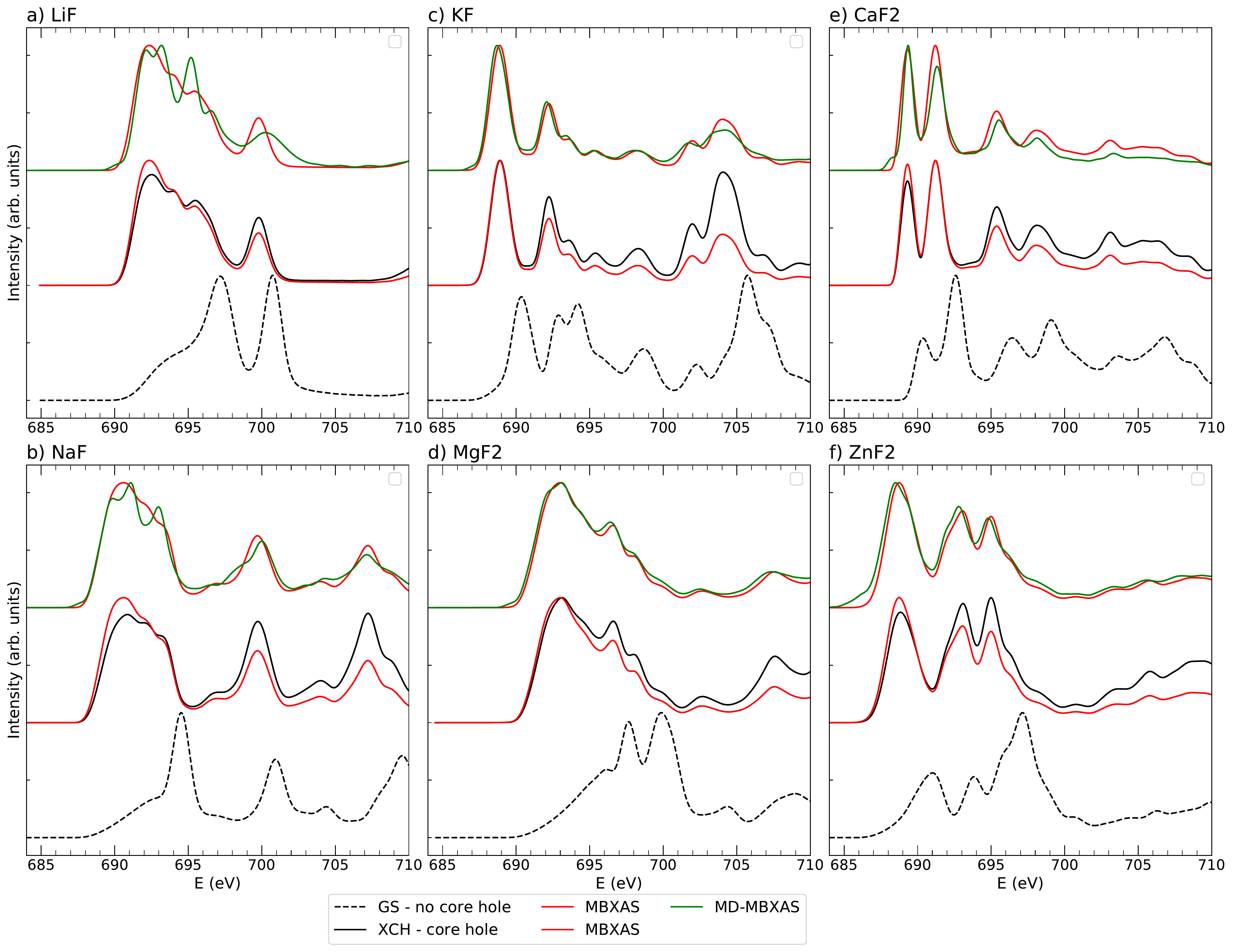}
\caption{Summary figure containing spectra of all solids at different levels of theory (from bottom to top): without core-hole (dotted line), XCH (black line), MBXAS (red line), and the MBXAS spectra including finite temperature effects (green line). Note that the supercell size for the MD-MBXAS calculations was 2x2x2 for all solids.}
\label{fig:sicalcall}
\end{figure*}

\bigskip

\section{Section 2: Effect of the level of theory on E$_{1s}$}
\label{sec:si_e1s}

First, we discarded numerical convergence issues (kinetic energy cutoffs, k-point sampling) and supercell size.  Then, we investigated the role of density functional choice, as the use of a semi-local functional like PBE can lead to overestimations of electronic covalency or delocalization. This is borne out by the apparently more accurate empirical estimates provided by Siegbahn's electrostatic expression, which is purely ionic (see above). Although introducing exact exchange increased the absolute electron binding energies for the solids, the relative values remained very similar. This very small effect even makes the binding energy of CaF$_2$ with respect to LiF larger, opposite than expected.  Another formulation of this hypothesis points to an overestimation of the polarizability of the metal, as observed 3$d$ mixing with the metal 3$p$ character in the 3rd row fluorides. Since ZnF$_2$ has a full $d$ shell, we tested the addition of a +U of 5 eV to  Zn, but the E$_{1s}$ remained essentially unchanged (0.03 eV).
\begin{table}[h!]
    \centering
    \begin{tabular}{c|c|c|c|c|c|c|c|}
    
     Solid & E$_{1s}$ & KE cutoff & k-points & Size & PBE0 & vdw-DF2 & 3s3pPP\\
     \hline
     LiF     &  -5.09  &  -5.08     & -5.09  & -5.29 & -6.18 & -5.54 & -5.00 \\
     CaF2    &  -1.40  &  -1.40     & -1.40  & -1.67 & -2.44 & -1.81 &-1.34\\ 
     \hline
     $\Delta$ & 3.69   & 3.68      & 3.69    &  3.62 & 3.74  & 3.73 & 3.66 \\
     \hline
    \end{tabular}
    \caption{Variation of E$_{1s}$ with kinetic energy cutoffs (40./320 Ry); k-point grid (5x5x5), supercell size (4x4x4 and 3x3x3 for LiF and CaF$_2$) and exchange-correlation functional (PBE0, vdw-DF2). }
    \label{tab:fch_testng}
\end{table}

\begin{figure*}[ht]
\centering
\includegraphics[width=1.\linewidth]{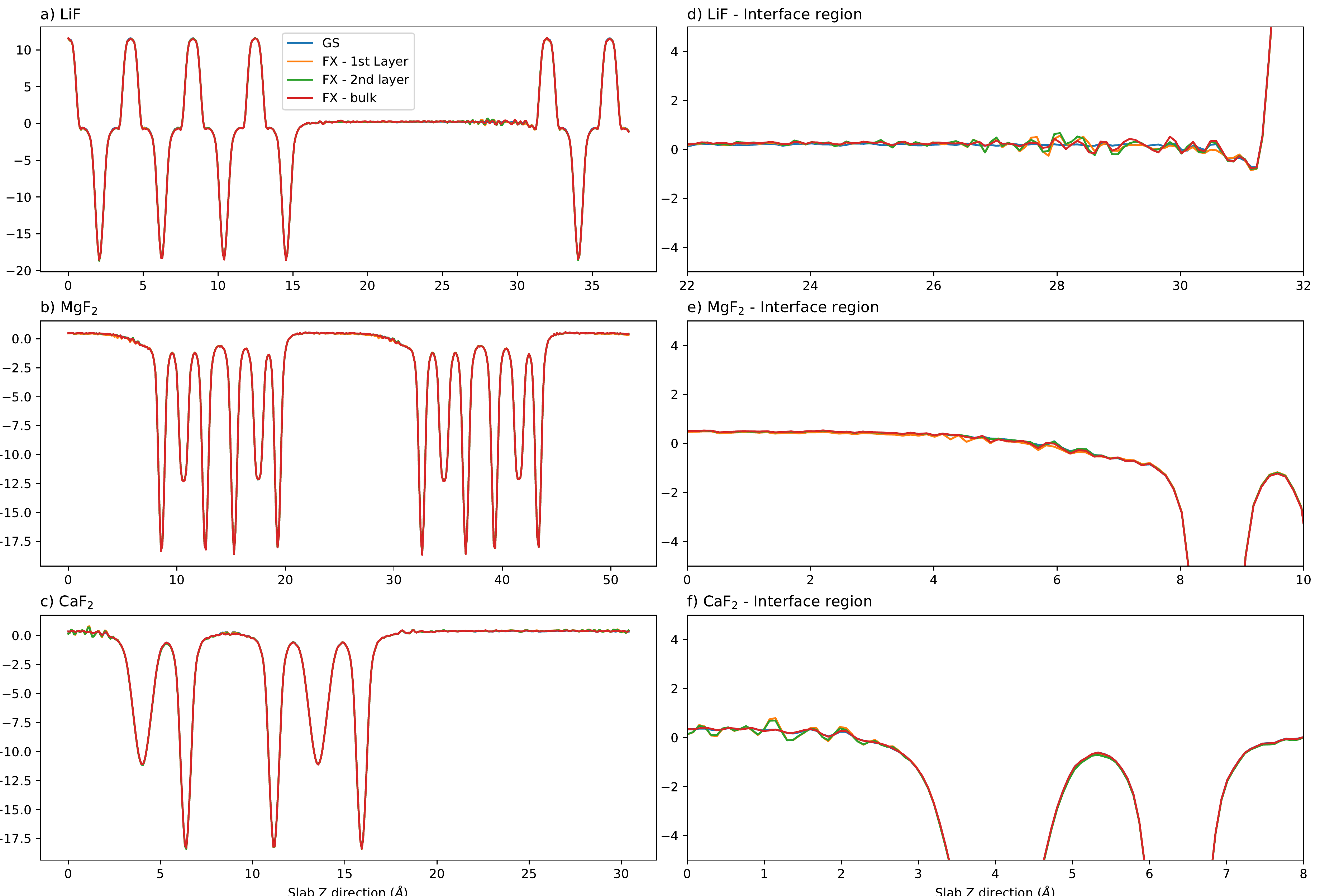}
\caption{Electrostatic potential in the direction perpendicular to the slab for the LiF(001), MgF$_2$(110) and CaF$_2$ surfaces (a - c) , the magnified interface region is shown on the right (d - f). The electrostatic potential reaches a plateau in the vaccuum region both in the ground state (red) and when the core-excited atom (FX) is placed in the bulk or outermost slab layers.  }
\label{fig:sislab_pot}
\end{figure*}

\begin{figure*}[ht]
\centering
\includegraphics[width=1.\linewidth]{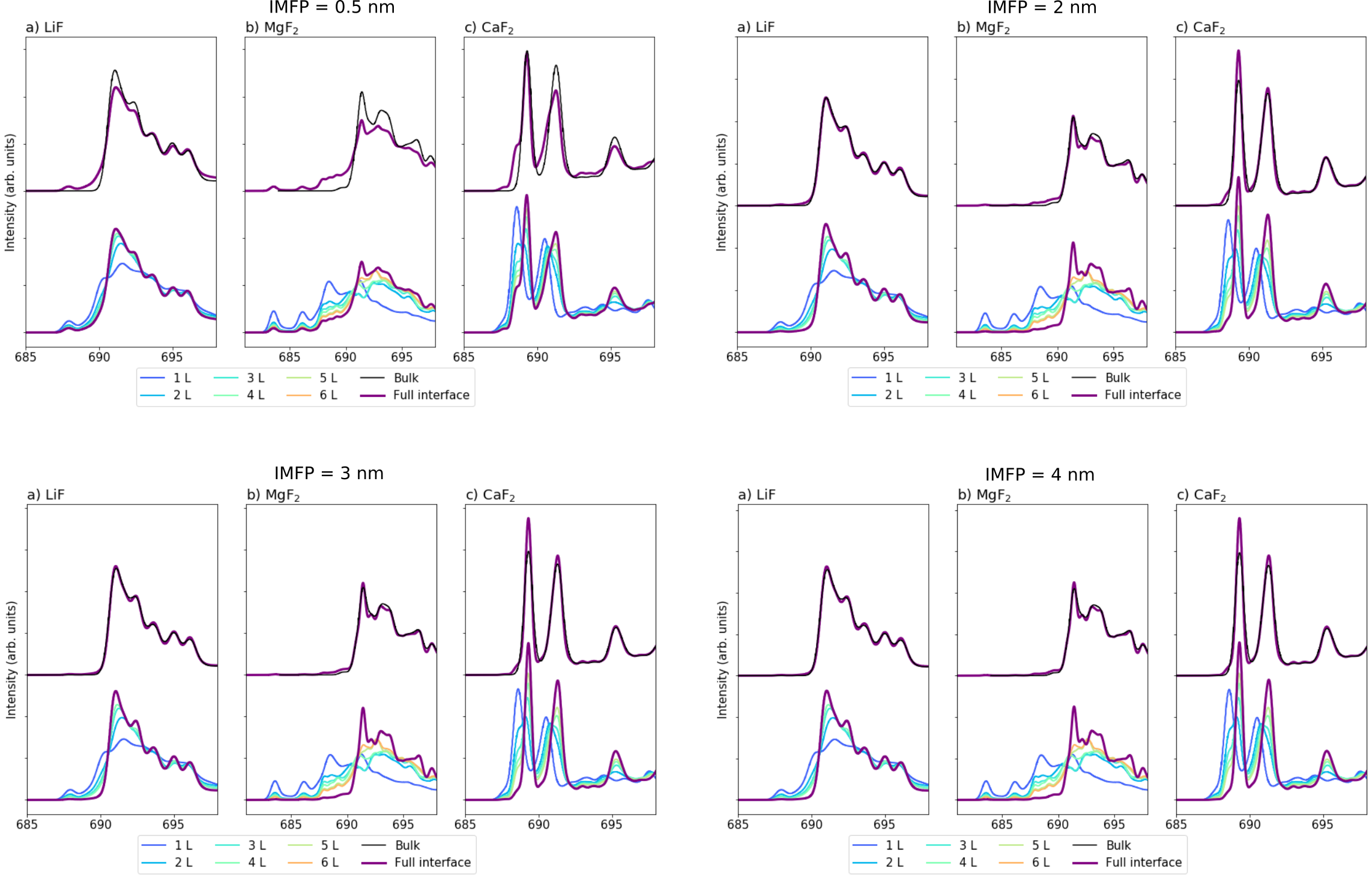}
\caption{Interfacial spectra at increasing IMFP values (0.5, 2, 3 and 4 nm). Although the intensity of the pre-edge is much higher at shorter IMFP values, it is still noticeable at IMFP=4nm. }
\label{fig:siinterface_imfp}
\end{figure*}

\begin{table*}[h!]
    \centering
    \begin{tabular}{c | c | c | c }
         Layer &LiF  & MgF$_2$ &  CaF$_2$   \\
         \hline
    1 &11 &5 & 9\\
    2& 10& 9 & 8\\
    3& 9& 4& 7\\
    4& 8& 4& 7\\
    5& -& 7& 6  \\
    6& -& 3& -\\
    bulk&62 & 68&63 \\
        \hline

    \end{tabular}
    \caption{Percentage of stoichiometrically-corrected contribution of each layer to the estimated interfacial spectra. }
    \label{tab:siweights_interface}
\end{table*}